\documentclass[prb,a4paper,twocolumn]{revtex4-2}
\usepackage{amssymb}
\usepackage{amsmath}
\usepackage{bbm}
\usepackage{amsfonts}
\usepackage{graphicx,epsfig,psfrag,xcolor,subcaption}
\usepackage{bm,comment}
\usepackage{color}
\usepackage[english]{babel}

\usepackage{float}
\graphicspath{{figures/}}

\def\ve{\varepsilon}

\def\vf{\varphi}

\def\arg{{\rm arg}}

\newcommand{\ignore}[1]{} 
\newcommand{\figref}[1]{Fig.~\ref{#1}}

\begin{document}
\title{Semiclassical scattering by edge imperfections in topological insulators under magnetic field}

 \author{A.~S.~Dotdaev}
  \affiliation{National University of Science and Technology MISIS, Moscow, 119049 Russia}
  \author{Ya.~I.~Rodionov}
  \affiliation{Institute for Theoretical and Applied Electrodynamics, Moscow, 125412 Russia}
  \author{A.~V.~Rozhkov}
    \affiliation{Institute for Theoretical and Applied Electrodynamics, Moscow, 125412 Russia}
  \author{P.~D.~Grigoriev}
  \affiliation{Landau Institute for Theoretical Physics, Russia}
 
\begin{abstract}
We study the scattering of edge states of 2D topological insulator (TI) in the uniform external magnetic field due to edge imperfections, common in realistic 2D TI samples. The external magnetic field breaks time reversal (TR) symmetry, opening the possibility of the scattering of otherwise topologically protected fermionic edge states.  The scattering happens to be always an over-barrier event, irrespective of the shape of the edge deformation and magnitude of the magnetic field.   
We use the advanced Pokrovsky-Khalatnikov semiclassical approach,  which allows us to obtain analytically both the main exponential and pre-exponential factors of the scattering amplitude for  wide classes of analytic deformation profiles.
\end{abstract}
\maketitle
\section{Introduction}
Topological insulators (TI) are new states of quantum matter which cannot be adiabatically transformed into
conventional insulators and semiconductors. They are characterized by a full insulating gap in the
bulk and gapless edge or surface states which are protected by time-reversal symmetry. The possible applications of TI include low-power electronics\cite{zutic2004} and error-tolerant quantum computing\cite{nayak2008,moore2009}. The properties of TI motivated considerable interest of scientific community since the transport by edge states in HgTe quantum wells (QW)\cite{konig2007} and surface states \cite{hsieh2008} in $\text{Bi}_2\text{Se}_3$ crystals had been experimentally observed.  The edge states are either 1D states on the boundaries of 2D TI (e.g. HgTe quamtum well) or 2D states on the boundaries of 3D TI (e.g. $\text{Bi}_2\text{Se}_3$ [\onlinecite{zhang2009}]). In most cases 2D and 3D TI samples are made from different compounds, the only exception being HgTe \cite{kvon2020}. Other realizations of 1D topologically protected states include those on the edges between surfaces of 3D TI \cite{deb2014} and states that occur on step edges \cite{herath2013,fedotov2017}.

A multitude of theoretical explanations \cite{hsu2021,yevtushenko2022} for the scattering mechanism of chiral edge (and surface) states have been developed in the last decade. They include the scattering from random fluctuations of the band gap \cite{tkachov2011}, the elastic scattering in charge puddles close to the edges \cite{vayrynen2014,essert2015,kurilovich2019prb,kurilovich2019prl}, the scattering from edge irregularities described by a certain coupling function \cite{magarill2015} and the scattering of counter-propagating states \cite{li2017}. The works studying the influence on conductance of the defects of certain kinds include precise analytical calculation \cite{zhangting2012,herath2013}, calculation of the reflectance from a single point-defect\cite{lupke2017}.

The main peculiarity of edge states in 2D TI is that due to spin-momentum locking, the scattering event (which is always a back-scattering in case of the edge of 2D TI)  always entails the flip of spin of a quasiparticle.
Therefore, in the absence of magnetic impurities, the elastic scattering of the edge states is strictly forbidden. This is the celebrated manifestation of TR symmetry of such systems~\cite{Kane2005}.  The important distinction of all TIs is their pronounced spin-orbit interaction (SOI),~\cite{hinz2006, yang2014}. In this paper we propose the model edge Hamiltonian describing the influence of SOI on edge imperfections. 
The edge imperfection is controlled by the deformation angle profile (see Fig. \ref{fig:bend}). The elastic scattering becomes possible in the presence of the uniform magnetic field orthogonal to the edge.

We build a comprehensive theory of scattering in such a system for wide classes of edge deformation profiles. A particular focus is put on the analytical structure of the solutions of the respective Dirac equation. We study the scattering in two complementary limits: the semiclassical limit of smooth edge deformation and perturbative limit corresponding to a small external magnetic field. The two limits are matched at the intersection of control parameters.

Our study allows to shed some light on how the intrinsic TR symmetry dictates the analytic properties of a scattering amplitude of the problem. In the absence of magnetic field we managed to find the exact solution for any  profile of the deformation potential. 
For the smooth deformation profiles the poweful Pokrovsky-Khalatnikov method~\cite{pokr1961} is used to obtain the analytic reflection amplitude with pre-exponential accuracy.

 The paper is organized as follows.
Section~\ref{model} is dedicated to the initial model and main approximations of the problem,
Section~\ref{methods} explains the main points of semiclassical Pokrovsky-Khalatnikov procedure and its application to the two main classes of scattering potentials, Section~\ref{born} discusses the exact solution of the magnetic-field-free problem and presents the perturbation (in magnetic field) theory, Section~\ref{match} discusses the matching of perturbative and semiclassical limit, we summarize the results in Section~\ref{discussion}.
 
\section{Model}
\label{model}

\begin{figure}[t!]
\label{fig:bend}
	\centering
	\includegraphics[width=0.8\columnwidth]{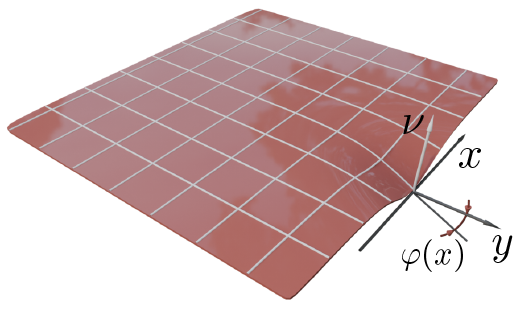}
	\caption{A schematic illustration of a geometric imperfection on the edge of a 2D topological insulator sample}
\label{fig:bend}\end{figure}

We assume the following form of Hamiltonian for the electrons in a 2D TI:
\begin{gather}
\label{eq:h1}
\hat{H} = h(\vec p)+\alpha\vec\sigma\times\vec p\cdot\vec\nu
\end{gather}
where $h(p)$ is the part of Hamiltonian defined by particular energy band structure, that is independent of spin-orbit interaction, and the spin-orbital interaction part is written in the same way as for 2D electron gas\cite{bychkov1984}; $\vec p$ is the momentum of electrons, $\nu$ is a unit vector perpendicular to the surface (or to an interface in a heterostructure), and $\alpha$ is Rashba parameter, which depends on the material and the external electric field (the gate voltage)\cite{hinz2006,yang2014}. The latter causes the splitting in energy bands due to electron's spin (Rashba splitting), which is vivid in energy band structure of TI materials [\onlinecite{zhang2010,yang2014}]. 

Here, it is important to mention how the direction of the normal vector $\nu$ in Eq.~\ref{eq:h1} should be chosen. In principle, the correct direction can be deduced from the initial TI Hamiltonian which we does not present here. However, we rely on the results of paper~\cite{zhang2010} where it is shown that the Fermi velocity of a TI grows as Rashba's coefficient $\alpha$ is reduced. As we are going to see below, the correct direction of $\nu$ corresponds to the one presented in Fig.~\ref{fig:bend}.
The energy bands of topological insulator materials have a linear part - Dirac cone, which is referred to as massless edge states. The expansion of the Hamiltonian near the Dirac point $p=p_0$ up to the first order in momentum gives:
$\pm v_0(p-p_0)\pm\alpha (p-p_0) = 
\pm (v_0-\alpha)|p-p_0|$,
where the sign in front of $\alpha$ depends on spin polarization, the sign in front of $v_0$ corresponds to the branches of energy bands; the parameter 
$v_0$ is interpreted as `bare' Fermi velocity. The momentum in~\eqref{eq:h1} has single component only ($x$ in our case) due to 1D nature of the edge electron transport. The effective Hamiltonian of edge states moving along x-axis (y=0) is  \cite{qi2011}
\begin{gather}
\hat{H}_0=v_F\hat{p}_x\hat\sigma_y.
\end{gather}
Here, $v_F=v_0-\alpha$ is Fermi velocity of the edge fermions renormalized by the spin-orbit interaction, $\vec p=(p_x,0,0)$ is the momentum of the fermions and $\sigma_{y}$ is the Pauli matrix acting in the spin 1/2 basis.
For the edge states in an ideal sample $H_{so}=\alpha\sigma_y p_x\equiv\Delta_{so}\sigma_y$, $\Delta_{so}$ is Rashba splitting, and there are no fluctuations of spin-orbit interaction.

Now consider a deformation located on the edge, like shown in \figref{fig:bend}. The edge tangent profile of the sample bend in $yz$ plane is defined by the function $\phi(x)$.
This defect leads to a new term of spin-orbit interaction: 
\begin{gather}
\alpha\vec\sigma\times\vec p\cdot\vec\nu
=-\alpha\hat p_x\hat\sigma_y + \alpha\hat p_x\hat\sigma_z\sin\phi(x)
\end{gather}
For smooth and shallow deformations it approximately equals
$ -\alpha \hat p_x\hat\sigma_y
+\alpha \hat p_x\phi(x)\hat\sigma_z$. The first term, $-\alpha\hat p_x\hat\sigma_y$ simply renormalizes Fermi velocity. The latter term, $\alpha p_x\phi(x)\hat\sigma_z$, is supposed to be treated as a perturbation, however it is not Hermitian. We consider the following symmetrization as perturbation:
\begin{gather}
\hat U(x)=\frac12\alpha(\hat p_x\phi(x)+\phi(x)\hat p_x)\hat\sigma_z
\end{gather}
In the following it would be convenient to include the parameter $\alpha$ in the profile function: $\vf=\alpha\phi$. 

The perturbation $\hat U$ alone will not disrupt ballistic transport of the edge states, since it does not break time-reversal symmetry. However, that is not the case in the presence of magnetic field, as we shall see. Let us apply magnetic field in the direction of $z$-axis (perpendicular to the plane of the topological insulator sample). We choose the following gauge of vector-potential: $\vec A=(\mathcal{H}y,0,0)$. The $y$ coordinate remains constant in our case $y=\rm const$ which can be safely put equal to zero (alternatively, for constant $y$ the vector potential can be gauged out from Dirac's equation with trivial gauge transformation). This way, the only change of the effective Hamiltonian is the addition of the interaction of the spin with the magnetic field (Zeeman term):
\begin{gather}
\label{ham1}
\hat{H}^{1D}(x)=v_F\hat{p}_x\sigma_y+\mu\sigma_z+\hat U(x),
\end{gather}
where $\mu=\mu_B g \mathcal{H}$, $g$ is g-factor for edge electrons \cite{kernreiter2016}. We applied transverse magnetic field because in-plane magnetic field has no effect on edge states, since the corresponding term in the Hamiltonian can be eliminated by a gauge transformation of the electron field operators \cite{zyuzin2011}.

Therefore, we need to solve the scattering problem for the following Dirac equation:
\begin{gather}
\label{dirac:main}
\left[v_F\hat{p}_x\sigma_y+\mu\sigma_z+\hat U(x)\right]\psi(x) = \ve\psi(x)
\end{gather}
It is easy to see, that even in the absence of deformation potential $\hat{U}(x)$ the Zeeman term $\mu\sigma_z$ opens a gap in the spectrum of edge state of the width $\mu$. 
Therefore, the propagated states always have the energy greater than $\mu$ and the condition:
\begin{gather}
    \label{zeeman}
    \ve>\mu.
\end{gather}
is always satisfied.

\section{methods}
\label{methods}
The Dirac equation
~\eqref{dirac:main} represents the system of two first order differential equations on the doublet $\psi = (\psi_1,\ \psi_2)$.
The most straightforward and (unexpectedly) convenient approach to its analysis happens to be the reduction of system~\eqref{dirac:main} to a 2nd order differential equation on a single function $\psi_1$. 

Despite a seemingly innocent look of Hamiltonian~\eqref{ham1}, the resultant equation has quite a beastly appearance. Since it is the central pillar of our work, we present it in the main body of the paper (this way the reader may also appreciate the magnitude of the disaster).
\begin{widetext}
\begin{align}
\label{main}
&2\hbar ^2 \left(\vf^2+1\right)\alpha \psi_1''  + 2 i \hbar  \left[\hbar ^2 \left(\vf^2+1\right) \vf ''+\vf \alpha \left(2 \mu -3 i \hbar  \vf '\right) \right] \psi_1'+ \left[\frac{1}{2}\alpha\beta(\alpha-2i\hbar\vf^\prime)+4\ve\hbar^2\vf\vf''\right]\psi_1=0,\\
\label{psi2}
 &\psi_2 = \frac{2 \hbar  \left(\vf ^2+1\right) \psi_1'-i\psi_1 \vf\beta(x)}{\alpha(x)},
\end{align}
\end{widetext}
where $\alpha(x) = 2 (\mu +\ve )-i \hbar  \vf ',\  \beta(x) = 2 (\mu - \ve )+i \hbar  \vf '$. 

The derivation of~\eqref{main} is straightforward, we present intermediate formulae for the reader's convenience in Appendix A.
We will call this equation \textit{Dirac equation} in what follows.
For obvious reasons, differential equation~\eqref{main} cannot be solved exactly. We will approach it from two distinct limits:\\ 
(i) semiclassical treatment, corresponding to the smooth deformation $\vf(x)$ of the edge;\\
(ii) perturbation theory in magnetic field strength (Zeeman energy) $\mu$.
\\
Then we will show how these two approaches beautifully match. 
\subsection{Semiclassical approximation}

The study of Eq.~\ref{dirac:main} in the semiclassical paradigm requires the stipulation of the small parameter of the analysis. In terms of physics, the semiclassical approach implies the smoothness of the system potential. 

In our case, the role of the potential is assumed by the edge deformation profile $\vf(x)$. The corresponding scale at which the potential changes is denoted as $a$.
Therefore, the smoothness of the potential means the de Broglie wavelength $\hbar v_F/\ve$ is much smaller than $a$:
\begin{gather}
\label{semi_cond}
    \frac{\lambdabar}{a}\equiv\frac{\hbar v_F}{\ve a}\ll1\quad (semiclassical\ approximation)
\end{gather}

As we will see in the following analysis, the structure of the semiclassical scattering in the problem is such that in view of condition~\eqref{zeeman} the semiclassical momentum never vanishes on the real axis, making the scattering  an \textit{over barrier } event. As is known from quantum mechanics, the latter entails the validity of semiclassical approximation  on the entire real axis~\cite{LandauIII}. One thus cannot recover the reflected wave in the WKB approximation in terms of perturbative  correction to the main semiclassical incident wave, due to the non perturbative (in small parameter~\eqref{semi_cond}) nature of the scattering amplitude~\cite{pokrovskii1958} (The reflected wave $\sim e^{-\#  a/\lambdabar}$). 
In what follows we put mostly $\hbar = v_F = 1$ for convenience, restoring them wherever necessary.

\subsection{Pokrovsky-Khalatnikov procedure}
The most powerful method to catch the reflected wave with pre-exponential accuracy in the over barrier scattering case is the Pokrovsky-Khalatnikov~\cite{pokr1961} (P-Kh) technique (see also an elegant work by M. Berry~\cite{berry1982semiclassically}). 
\begin{figure}[t]
	\includegraphics[width=0.5\columnwidth]{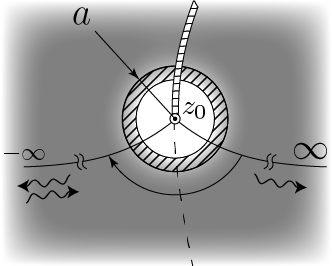}
\captionsetup{justification   =  raggedright,
              singlelinecheck = false}
	\caption{Towards Pokrovsky-Khalatnikov method. The vicinity of the turning point $z_0$ with a customary branch cut extending upward. The gray area denotes the range of validity of the semiclassical approximation (extending to infinity in the complex plane). The circle of radius $a$ denotes the range of applicability of the exact solution near the turning point. The two solid curves sprawling from the turning point to $\pm\infty$ are the anti-Stokes lines. The dashed line is the Stokes line. The striped region is the region where both the semiclassical and exact solutions are valid and can be matched.}
\label{fig:pokrov1}
\end{figure}
The idea can be summarized in the following steps:

(i) Perform the analytical continuation of the semiclassical solution into the complex plane along a so called anti-Stokes line, ${\rm Im}\int_{z_0}^z k(z)\,dz= 0$ where $k(z)$ is the semiclassical momentum and $z_0$ is the turning point in the complex plane. The continuation along the anti-Stokes lines is vital, since both solutions $\exp(\pm i\int kdz)$ have constant modulus on this line and can be easily distinguished from one another 
(see Fig.~\ref{fig:pokrov1}, anti-Stokes lines go to the left and right from point $z_0$).

(ii) Build the exact solution of the Schroedinger equation in the vicinity of the turning point $z_0$ (where the momentum $k(z)$ can be Taylor-expanded and the equation drastically simplifies). 

(iii) Find the asymptotics of the exact solution on the anti Stokes lines going to the right and to the left from the turning point.

(iv) Assuming, there is a non-empty intersection of the range of validity of the asymptotics of exact solution and semiclassical solution (the striped region in Fig.~\ref{fig:pokrov1}) match the semiclassical and exact solutions in the mentioned range on anti-Stokes lines.  
(The problem can be solved in the semiclassical approach if and only if such an intersection exists)

(v) Build an analytic continuation from the anti-Stokes line going to $-\infty$ on the real axis $\psi(z)\rightarrow\psi(x)$.

To understand the last line of point (i) better, one is advised to think of the behavior of both solutions on the Stokes line  ${\rm Re}\int_{z_0}^z k(z)\,dz= 0$ (dashed line in Fig.~\ref{fig:pokrov1}), which is the steepest descent line for the real part of the exponent of both solutions.
The Stokes line is, therefore, the line of maximal domination of the one solution over the other, where the suppressed solution gets smaller than the error of the dominating solution.

In what follows we are going to implement the outlined program step by step explaining all the nuances.

\subsection{Semiclassical solution}

We introduce the exponential substitute $\psi\ \rightarrow e^{iS/\hbar}$ for the wave function and employ the standard semiclassical machinery adapted to the stationary Dirac equation:
\begin{gather}
\label{psiseries}
\psi=\begin{pmatrix} \psi_1 \\ \psi_2 \end{pmatrix}
,\quad
\psi_{1,2}=\exp\left(\frac{iS_0}{\hbar}+i S_{1,2}+...\right),
\end{gather}
where the semiclassical action is expanded as series with respect to $\hbar \rightarrow 0$ (in powers of $\hbar v_F/(\ve a)$ to be precise).
In the zeroth order in $\hbar$ (or discarding all terms with derivatives of $\vf$ in Eq.~\ref{main}) we obtain the following expression for $S_0$ (Appendix A):
\begin{align}
\label{action}
    S_0(x)&=\int\limits^{x} q_{\pm}(x')\,dx',\\
\label{mom_semi}
    q_{\pm} &= \frac{-\mu  \vf\pm p}{\vf^2+1},\quad
   p=\sqrt{\ve ^2(\vf^2+1)-\mu ^2}
\end{align}
where $q_\pm$ is interpreted as semiclassical momentum. The regular branch of $p$ is chosen in such a way that $p\underset{x\rightarrow+\infty}{\rightarrow}\sqrt{\ve^2-\mu^2}$.
Then, retaining the next terms of order $\hbar$ ($S_{1,2}$ in the substitute \eqref{psiseries}), and plugging it back in~\eqref{main} we obtain the pre-exponential semiclassical terms for the wave function $\psi$:
\begin{gather}
\label{hbar1}
    \begin{split}
        \psi_{1,\pm}(x) &= \xi_{1,\pm}(x)\exp\left(\frac{i}{\hbar}\int q_\pm\,dx\right)\\
        \xi_{1,\pm}&=\sqrt{\pm q_{\pm}\Big[1\pm\frac{\vf\ve}{p}\Big]}\\
        \psi_{2,\pm}(x) &=-i\psi_{1,\pm}\frac{\ve\vf\mp p}{\ve+\mu}
    \end{split}
\end{gather}
Again,  the external square roots entering the definition of $\xi_{1,\pm}$ are assumed to be positive at $x\rightarrow+\infty$. 
To clear out which of the solutions corresponds to the right (left) moving carriers we need
semiclassical currents:
\begin{gather}
    \begin{split}
        j_{\pm} &\equiv \psi_{\pm}^\dag \sigma_y \psi_{\pm} =\frac{2q_{\pm}}{p}(\ve-\mu)
    \end{split}
\end{gather}
At $x\rightarrow\pm\infty$ the profile function $\vf(x)\rightarrow 0$. Therefore, 
\begin{gather}
    j_\pm \underset{x\rightarrow\infty}{=}\pm2(\ve-\mu)
\end{gather}

\subsection{The study of the turning points}
Now we need to address the points where the semiclassical approach breaks down. Those are usually the branch points of the semiclassical momentum $q_\pm$~\eqref{mom_semi}. These points hint at the possible singular points~\cite{WW} of the differential equation, where semiclassical approach fails. 

\subsubsection{The branch points of $p$.}
These are simultaneously the branch points of $q_\pm$:
\begin{gather}
    \label{branch1}
    \vf(z_\pm)\equiv\vf_{\pm}= \pm i\sqrt{1-\frac{\mu^2}{\ve^2}}\\
    \label{branch2}
    \vf(z)=\vf_\pm+(z-z_\pm)/a+...
\end{gather}
and semiclassical momenta corresponding to different linear independent solutions cannot be distinguished. We assume that the branch points are of the simplest structure, meaning they are the first order roots of $p^2(z)$. 
The corresponding branch point  is always complex due to condition $\ve>\mu$ (see expression ~\eqref{semi_cond}). 

Expanding the potential $\vf(z)$ in the vicinity of the points $z_\pm$, (Eq.\ref{branch2}), we immediately arrive at the semiclassical condition~\eqref{semi_cond} (after some simple but cumbersome algebra, see \ref{app:branch1}) in the form:
\begin{gather}
\label{semi1}
    \frac{\ve\hbar^2 v_F^2 }{(\ve^2-\mu^2)^{3/2}|a||z-z_\pm|}\ll 1
\end{gather}
which breaks at $z\rightarrow z_\pm$.
Therefore, points $z_\pm$ should be included into the implementation of P-Kh method. 

Here, we notice that parameter $a$ in the expansion~\eqref{branch2} is, in principle, complex. However, its modulus $|a|$ corresponds to the characteristic length scale of the potential $\vf(x)$. 

\subsubsection{The zeroes of $q_{\pm}$.}
These points, if they exist, are the most natural candidates for the breakdown of the semiclassical approach.
They obviously correspond to the singularity of the edge profile $\vf(z)$. We assume the simplest, yet the most common type of singularity, namely, a simple pole:
\begin{gather}
\label{laurent}
    \vf(z) = \frac{ia}{z-z_p}+...
\end{gather}
Here, as before $|a|$ corresponds to the characteristic length of the potential. The imaginary unity $i$ is introduced for convenience. In the same manner as was obtained condition~\eqref{semi1} we procure the semiclassical condition in the vicinity of the turning point $z_p$,  $q_\pm(z_p)=0$:
\begin{gather}
\label{semi2}
\frac{\hbar v_F a\ve}{(\ve^2-\mu^2)|z-z_p|^2}\ll1
\end{gather}
As expected, condition~\eqref{semi2} is also broken as $z\rightarrow z_0$. Therefore,  $z_p$ should also be included in the P-Kh procedure. 
\subsubsection{The relative location of $z_p$ and $z_\pm$.}
In general situation, when the amplitude of the potential on the real axis is $\sim 1$, the only parameter of the potential is its length scale $a$. Therefore, we conclude that $|z_p|\sim |a|$ in~\eqref{laurent}. 
Then, the convergence radius of the Laurent series~\eqref{laurent} is again of the order of $a$. At $|z-z_p|\sim a$ the potential has the value $\vf\sim 1$.  
On the other hand, at points $z_\pm$ the potential has the values $\vf_\pm$ (see Eq.~\ref{branch1}) which are also $\sim 1$ (unless $\ve-\mu\ll\ve$ which is not considered in this paper).

Therefore, qualitatively, we expect that the branch points $z_\pm$ belong to the convergence circle of the Laurent expansion ~\eqref{laurent}. Thus we can approximately locate the branch points $z_\pm$ via the equation:
\begin{gather}
    \frac{i a}{z-z_p}\approx \pm i\sqrt{1-\frac{\mu^2}{\ve^2}}
\end{gather}

As a result, we have the approximate typical relative position of the branch points of the semiclassical momentum $q$ and the poles of the deformation field $\vf$. 
\begin{gather}
    z_{\pm}=z_p\pm\frac{a}{\sqrt{1-\frac{\mu^2}{\ve^2}}}.
\end{gather}
Therefore, the branch points $z_\pm$ and the pole $z_p$ are always separated by the distance of the order of the length scale $|a|$ of the potential.  

\subsubsection{On the analytical structure of $\vf(z)$.}

Here, we should recall the following fact from complex analysis. If the deformation profile $\vf(z)$ is not a constant function, it must have singularities in 
the extended complex plane. Therefore, function $\vf(z)$ is split into general two classes: the one which possess singularities at finite points (the typical example would be the Lorentzian potentials), and the one which has 
no singularities at any finite point (e.g. the Gauss-type-potentials). In the latter case, the following remark is in order. Since $\vf(x)\rightarrow 0,\ x\rightarrow\pm\infty$ the latter class corresponds to the situation where $\vf(z)$ 
has an essential singularity at $z\rightarrow\infty$. The typical example would be $\vf(z)=P_n(z/a)e^{-z^2/a^2}$, where $P_n(z/a)$ is a polynomial of the order 
of $n$.
Our treatment of the problem is thus split into two cases:

(i) The case of regular at any finite point potential.

(ii) The case of potential with singularities at finite points, where the type of singularities is restricted to simple poles. 

\subsection{Regular potential}
\label{reg1}
This case corresponds to the classic P-Kh treatment adapted to a more complicated Dirac equation.
\subsubsection{Transformation from Dirac to Schroedinger equation}
To make the analogy between Dirac equation Eq.~\ref{main}  and Schrödinger equation more pronounced, we get rid of the first derivative in \eqref{main} via a standard substitute~\cite{WW}.
The equation is transformed as follows:
\begin{align}
    &\psi''(x)+\eta(x)\psi'(x)+\kappa(x)\psi(x)=0 \Rightarrow\notag\\
    \label{eq_mod}
    &\theta''(x)+\pi^2(x)\theta(x)=0\ \ {\rm (Schroedinger\ equation)}\\
    \label{eq_mod2a}
    &\theta(x) = \exp\left(\frac{1}{2}\int^x \eta(t)dt\right)\psi(x),\\
    &\pi^2(x) = \kappa(x)-\frac{1}{2}\eta'(x)-\frac{1}{4}\eta^2(x).
\end{align}
The expression for $\pi^2(x)$ is quite cumbersome. However, it is instructive to write down $\eta(x)$ and $\pi^2(x)$ discarding all the derivatives of the potential field $\vf(x)$ (zeroth semiclassical approximation) as well as semiclassical solution. This way the connection with the initial semiclassical relations ~\eqref{action}, \eqref{mom_semi} becomes transparent:
\begin{align}
\label{semi3}
        \eta(x)&=\frac{2i}{\hbar}\frac{\mu\vf(x)}{\vf^2(x)+1},\quad
        \pi^2(x)=\frac{\ve^2(\vf^2+1)-\mu^2}{(\vf^2+1)^2}\\
        \theta_\pm(x)&=\frac{1}{\sqrt{\pi(x)}}\exp\left(\pm i\int_{x_0}^x \pi(t)dt\right).
    \label{semi4}
\end{align}
In the last equation point $x_0$ needs to be chosen on the real axis. This way both functions $\theta_{\pm}$ have the same modulus. Apart from this $x_0$ is quite arbitrary and is picked from convenience considerations. 
To be on the safe side, we show in Appendix~\ref{cooresp} how~\eqref{semi3}-
\eqref{semi4}
exactly reproduce full semiclassical expressions ~\eqref{action}-\eqref{hbar1} following from the Dirac equation. 
\subsubsection{Exact solution near the turning point of $p$}
\label{subsec:branch}
The exposition now follows the celebrated work~\cite{pokr1961}. 
Still, we would like to present some details here, so the reader understands the machinery of the method to be able to follow the next sections which are more involved. 

It is time to implement step (ii) of the P-Kh method. For concreteness (and to make notation more concise), we need to make a choice which point is going to have the dominant contribution to the reflection coefficient. As usual (and it will be clearly seen later), it is the closest to the real axis point.
We assume it is the point $z_+$.
In the situation with model potentials, several branch points can be equidistant from the real axis (e.g. $\vf(z) = [\cosh(z/a)]^{-1}$). In this situation, all the points contribute equally to the reflection amplitude. Though attractive from the aesthetic viewpoint and resulting in beautiful answers with quantum oscillations,  these situations are unrealistic and will not be discussed here.

\begin{figure}[t]
	\includegraphics[width=0.85\columnwidth]{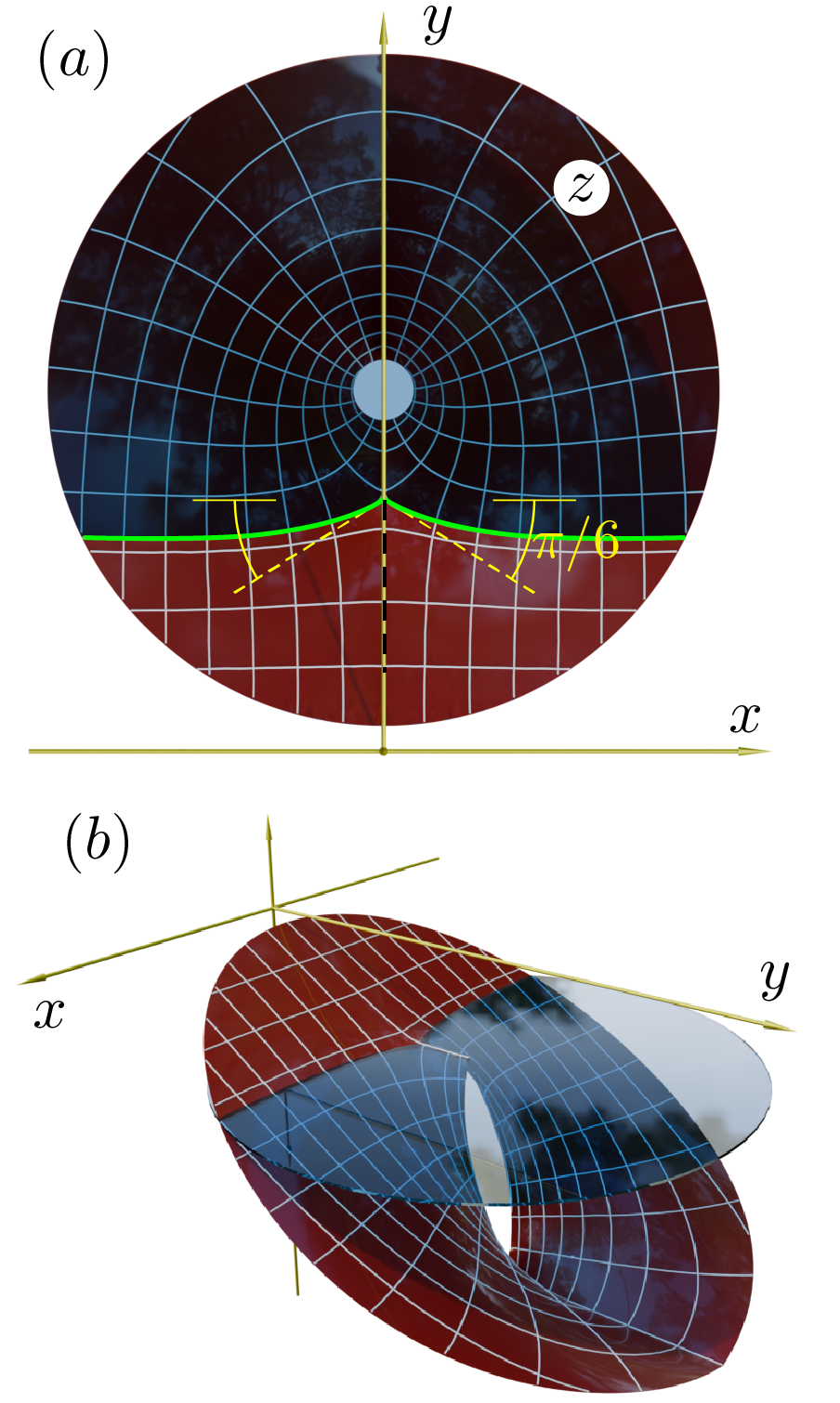}
\captionsetup{justification   =  raggedright,
              singlelinecheck = false}
	\caption{Exact anti-Stokes lines for the deformation profile $\vf(z)=ze^{-z^2/2}$ presented as a cross-section of the integral $h (z) = {\rm Im}\int_{0}^z\pi(t)\,dt$ and the level plane $h = h(z_+)$. (a) The top view. The anti-Stokes line marked as a green line. The Stokes line is presented as a black dashed line. (b) Axonometric view. One clearly sees, how the anti-Stokes line becomes horizontal far from the turning point. The hole in the center is the circular vicinity of the singularity $z_0$ of $\pi(z)$, where $\vf(z_0)=i$.}
\label{fig:anti-Stokes}
\end{figure}

 With the help of~\eqref{branch2} we expand the semiclassical momentum near the branch point:
 \begin{gather}
 \label{exp1}
     \pi^2(\zeta) = \frac{1}{\hbar^2}\frac{2i\zeta}{a}\frac{\epsilon ^5 \sqrt{\ve^2-\mu ^2}}{\mu ^4}+...,\quad \zeta = z-z_+
 \end{gather}
 so that equation ~\eqref{eq_mod} is reduced in the vicinity of the branch point   $z_+$
 to the classical Airy equation:
\begin{gather}
\label{eq-mod2}
\theta''(s)+s\theta(s)=0,\quad s=\gamma^{1/3}\zeta,\ \ \gamma = \frac{2i\ve^5}{a\hbar^2\mu^4}\sqrt{\ve^2-\mu^2}.
\end{gather}
See the details and the subtleties of the derivation of~\eqref{eq-mod2} in Appendix~\ref{appendix:semi2}.
The asymptotics of the Airy function at large value of the argument are well-known and should be matched with semiclassical expansion.  Before proceeding any further, we determine the position of the anti-Stokes lines using the semiclassical analysis of Eq.~\eqref{eq-mod2} at large values of $\zeta$: $\theta(\zeta)\propto \exp[i\int^\zeta \pi(t)dt],\ \ \int^\zeta \pi(t)dt\gg1$:
\begin{gather}
\label{anti-Stokes1}
    \begin{split}
    {\rm Im}\int\limits_0^s \pi(t)\,dt &=0 \Rightarrow\ 
    {\rm Im} s^{3/2}= 0\ \Rightarrow\\
    \arg s &= \frac{2\pi n}{3},\ \ n\in \mathbb{Z}
    \end{split}
\end{gather}
As we see from~\eqref{anti-Stokes1}, the anti-Stokes lines are separated by angle $2\pi/3$. 
The pattern of anti-Stokes lines for the typical profile function $\vf(z)=ze^{-z^2/2}$ is presented in Fig.~\ref{fig:anti-Stokes}, where the anti-Stokes lines are drawn as level lines of the surface of semiclassical action with exact momentum ${\rm Im}\int_0^\zeta\pi(t)dt$.
The anti-Stokes lines in Fig.~\ref{fig:anti-Stokes} are drawn for non-rescaled coordinate $z$.  

\begin{figure}[t]
	\includegraphics[width=1\columnwidth]{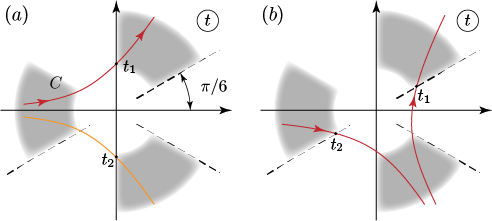}
\captionsetup{justification   =  raggedright,
              singlelinecheck = false}
	\caption{(a) The placement of the contour $C$ defining solution Eq.~\ref{exact1} for $\arg s = 0$. The asymptotic gray areas are the regions which should contain the contour end points. (for the integral to converge). $t_1,\ t_2$ are the saddles of the integrand. The contour drawn via exact steepest descent path passing through saddle $t1$, the orange line is the second steepest descent path passing through saddle $t_2$. (b) The steepest descent paths and the saddles for the $\arg s = -2\pi/3$ (left anti-Stokes line in Fig.~\ref{fig:anti-Stokes}).}
\label{fig:asymp2}
\end{figure}

As a result, the semiclassical wave functions ~\eqref{semi4} in the vicinity of $z_\pm$ on the anti-Stokes line $n=0$ (the one going to the right and corresponding to the outgoing wave) assume the following form:
\begin{gather}
    \label{asy1}
    \theta^{\rm app}_{\pm}(s) = \frac{1}{s^{1/4}\gamma^{1/6}}
    \exp\left(\pm\frac{2i}{3}s^{3/2}\right),\quad s\gg1
\end{gather}
The exact solution of the Airy equation~\eqref{eq-mod2} which has the asymptotics~\eqref{asy1} has the following integral representation:
\begin{gather}
  \label{exact1}
    \theta(s) = \frac{e^{-i\pi/4}}{\sqrt{\pi}\gamma^{1/6}}\int\limits_C e^{st+t^3/3}\,dt,
\end{gather}
where contour $C$ is presented in Fig~\ref{fig:asymp2}(a).  
To make our presentation self-contained, we derive integral representation~\eqref{exact1} in Appendix~\ref{app:airy}.
The contour is placed in such a manner that only saddle $t_1 = i\sqrt{s}$ contributes to the asymptotics.
The constant in front of the integral~\eqref{exact1} is chosen in such a way that the asymptotics of~\eqref{exact1} at $s\rightarrow\infty$ coincides with semiclassical expression~\eqref{asy1} for $\theta_+^{\rm app}(s)$ exactly. As we turn from right anti-Stokes line to the other (going to the left at angle $-2\pi/3$) the saddle points of the integrand in ~\eqref{exact1} rotate in the complex plane of $t$. Once we cross the Stokes line (see Fig.~\ref{fig:anti-Stokes}) the topology of the steepest descent paths changes and both saddles start to contribute to the asymptotics of~\eqref{exact1} (see Appendix for details). This way we have the solution for the anti-Stokes line sprawling to the left:
\begin{gather}
    \label{stokes-left}
    \theta_<(s) = \frac{1}{\gamma^{1/6}}\left(\frac{\exp\left[\frac{2i}{3}s^{3/2}\right]}{s^{1/4}}-i\frac{\exp\left[-\frac{2i}{3}s^{3/2}\right]}{s^{1/4}}\right).
\end{gather}

\subsubsection{Matching exact and semiclassical solutions}
We need to make sure that the region of validity of semiclassical expansion and exact solution have nonzero intersection. The validity of the semiclassical expansion ~\eqref{asy1} is $s\gg1\ \Rightarrow$ 
\begin{gather}
\label{region_semi}
    |z-z_+|\gg (a\hbar^2/\ve^2)^{1/3}.
\end{gather}
On the other hand, the validity of the Taylor expansion of potential profile is $|z-z_+|\ll a$. The equation of coexistence of the last two conditions is the smallness of parameter $\ve a/\hbar$, which is precisely the principal semiclassical expansion parameter of the problem (Eq.~\ref{semi_cond}).  

We see that the solution to the left of the barrier is split into two waves and perfectly matches with semiclassical expressions~\eqref{asy1} in the same region. Thus, we substitute both waves in~\eqref{stokes-left} with semiclassical waves~\eqref{semi4}:
\begin{gather}
\label{semi5}
    \theta_<(z)= \theta_+(z)\Big|_{z_+} - i\theta_-(z)\Big|_{z_+}.
\end{gather}
In the last equation, we return from $s$
to $z$. The integral defining the exponent in functions $\theta_\pm$ in ~\eqref{semi4} is assumed to have the lower integration limit changed from $x_0$ to $z_+$. 

Finally, we make an analytical continuation of the wave function~\eqref{semi5} on the real axis $z\rightarrow x$. The structure of the equation becomes especially transparent with the help of the relation:
\begin{gather}
    \theta_{\pm}(x)\Big|_{z_+} = \exp\left(\pm i\int\limits_{z_+}^{x_0}\pi(z)\,dz\right)\theta_{\pm}(x),
\end{gather}
where $\theta_{\pm}$ are defined in Eq.~\eqref{semi4}. Then we obtain the reflection amplitude:
\begin{gather}
\label{reflection1}
    r = \exp\left(-2i\int\limits_{z_+}^{x_0}\pi(z)\,dz\right)
\end{gather}
and the reflection coefficient:
\begin{gather}
\label{reflection2}
    R \equiv|r^2| =\exp\left(-\frac{4}{\hbar v_F}{\rm Im}\int\limits_{x_0}^{z_+}\frac{\sqrt{\ve^2(\vf^2+1)-\mu^2}}{\vf^2+1}\,dz\right),
\end{gather}
where we restored the Fermi velocity $v_F$ for convenience.
The reflection coefficient~\eqref{reflection2} is the first main result of our paper. 
It has a typical semiclassical appearance. The exponent of the exponential function can be estimated as $\#\ve a/\hbar\gg1$ making the reflection coefficient exponentially small. 

Despite its simplicity, result~\eqref{reflection2}
has some peculiarity. Namely, it is by no means obvious how to make a transition from arbitrary $\mu<\ve$ to $\mu\rightarrow0$ (vanishing external magnetic field) in formula~\eqref{reflection2}. 
The point is, at vanishing $\mu$ the TR symmetry of the problem is restored, and the reflection coefficient ought to vanish exactly. The modification of result ~\eqref{reflection2} reflecting this fundamental symmetry of our problem is going to be by far the most nontrivial
part of the paper, which will be discussed in the section after the study of Born approximation.



\subsection{The potential with a pole}
\label{pole}
\subsubsection{The exact solution in the vicinity of the turning point}
\begin{figure}[t!]
	\centering
\includegraphics[width=0.85\columnwidth]{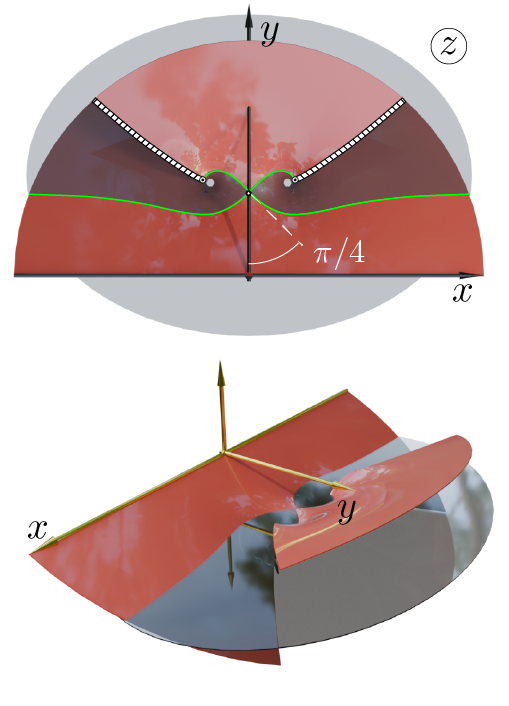}
	\caption{Exact anti-Stokes lines for the deformation profile $\vf(z)=(z^2+1)^{-1}$ presented as a cross-section of the integral $h (z) = {\rm Im}\int_{0}^z\pi(t)\,dt$ and the level planr $h = h(z_+)$ (glass surface). (a) Te top view. The anti-Stokes line marked as a green lines. We see that only lower anti-Stokes lines sprawl to $\pm\infty$ and become eventually parallel to the real axis. Upper anti-Stokes lines end at branch cuts origination from singular points $\ve^2(z)=-1$. (b) Axonometric view of the cross section of the surface with a level glass plane.}
\label{fig:stokes2a}
\end{figure}

The equation for anti-Stokes lines is easily obtained in the vicinity of point $z_p$ along the lines outlined in Section~\eqref{subsec:branch} (see Eq.~\ref{anti-Stokes1}). With the help of potential expansion~\eqref{laurent} we obtain: 
\begin{align}
    {\rm Im}\int_0^\zeta \pi(t)\,dt &=
    {\rm Im}\left[-\frac{i\ve}{a}\int_0^\zeta tdt\right] = 0\ \Rightarrow \notag\\
    \arg\zeta &= \frac{\arg a}{2}+\frac{\pi}{4}+\frac{\pi n }{2},\quad n\in\mathbb{Z}.
    \label{anti-Stokes2}
\end{align}
Here, as before $\zeta \equiv z-z_p$. 
We see that anti-Stokes lines form $\pi/4$ directions (up to the rotation by $\arg a$) with the real axis. As an example we draw the portrait of the lines for a Lorentzian type profile $\vf(z)=(z^2+1)^{-1}$ in Fig.~\ref{fig:stokes2a}.

The details of analytical continuation of the semiclassical wave functions~\eqref{hbar1} in the vicinity of a simple pole $z_p$ along the anti-Stokes lines~\eqref{anti-Stokes2} are given in Appendix~\ref{app:pole}. We get the following asymptotics:
\begin{gather}
\label{qctpasympt}
\begin{split}
      \psi_{1+,\gtrless}(\zeta)&= 
      \frac{\ve+\mu}{\ve}\sqrt{\frac{\ve-\mu}{2}}\left(\frac{\zeta}{a}\right)^{\frac{3}{2}} e^{-\frac{\ve+\mu}{2a}\zeta^2+\frac{3\pi i}{4}}
      \\
      \psi_{1-,\gtrless}(\zeta)
      &=
      \sqrt{2(\ve-\mu)}\left(\frac{\zeta}{a}\right)^{\frac{1}{2}} e^{\frac{\ve-\mu}{2a}\zeta^2+\frac{i\pi}{4}},
\end{split}
\end{gather}
where, as before, symbol $>(<)$ corresponds to the semiclassical solution on the right (left) anti-Stokes line in the vicinity of the turning point $z_p$. The relations~\eqref{qctpasympt} are written in the semiclassical limit $\zeta\gg\sqrt{|a|/\ve}$.

Now we need to get an exact solution of Dirac equation~\eqref{dirac:main} in the vicinity of the turning point $z_p$. However, the expansion of  equation~\eqref{dirac:main} at $z_p$ is rather cumbersome and we relegate the reader to Appendix~\ref{app:trans}. Fortunately, the educated substitute  
$\psi_1(\zeta)= \exp(\zeta^2/2a[\ve-\mu])\sqrt{\zeta}\psi(\zeta)$ leads to a disappearance (!) of the term without derivative in~\eqref{dirac:main}, giving rise to a much simpler differential equation:
\begin{gather}
\label{dif_simp}
   a  \psi ''\!\!\left[2 \zeta^2 (\mu\!+\!\ve )
   \!-\!a \right]\!+\!\psi '\!\!\left[4 \zeta^3 \ve  (\mu\!+\!\ve )\!-\!2 a  \zeta (2 \mu +3 \ve )\right]\!=\!0
\end{gather}
which is trivially integrated in quadratures:
\begin{gather}
\label{psi1exact}
\begin{split}
    \psi_1(\zeta) &=\sqrt{\frac{\zeta}{a}} \left(c_1-c_2\frac{\sqrt{\pi}\mu}{2\ve^{3/2}\sqrt{a}}\text{erf}\Big[\sqrt{\ve a}\frac{\zeta}{a}\Big]\right)e^{\frac{\ve-\mu}{2a}\zeta^2}\\
    &+c_2\frac{\ve+\mu}{\ve}\left(\frac{\zeta}{a}\right)^{3/2}e^{-\frac{\ve+\mu}{2a}\zeta^2}
\end{split}
\end{gather}
where erf is the error function, $\text{erf}(z)=\frac{2}{\sqrt{\pi}}\int\limits_0^z e^{-t^2}dt$.
Now we have all the ingredients to match exact solution~\eqref{psi1exact} with semiclassical wave functions~\eqref{qctpasympt}.
\subsubsection{Matching semiclassical and exact solution}
To get the reflection coefficient, we require that at the right anti-Stokes line there is only a transmitted semiclassical wave, meaning the asymptotics of solution ~\eqref{psi1exact} coincides with function $\psi_{1+,>}$ in~\eqref{qctpasympt}. This immediately yields the condition (see Appendix~\ref{app:erf} for the asymptotics of erf function):
\begin{gather}
\label{const1}
    c_1 = \frac{\sqrt{\pi}\mu}{2\ve^{3/2}\sqrt{a}}c_2,\quad c_2 = \sqrt{\frac{\ve-\mu}{2}}e^{3\pi i/4}.
\end{gather}
The expression for $c_2$ follows from the requirement of exact match with $\psi_{1+,>}$ in~\eqref{qctpasympt}.
This way, we have for the solution on the left anti-Stokes line:
\begin{gather}
\label{asymp_pole}
\psi_1(\zeta)\Big|_{\rm left}=\psi_{1+,<}(\zeta)+\frac{i\sqrt{\pi}\mu}{2\ve^{3/2}\sqrt{a}}\psi_{1-,<}(\zeta).
\end{gather}
Changing semiclassical expressions $\psi_{1\pm,<}$ to general expressions~\eqref{hbar1}, we are able to perform the analytical continuation from  the anti-Stokes line to the real axis $x$. Noticing that $\xi_{1,+} = \xi_{1,-}$ at $x\rightarrow-\infty$ in~\eqref{hbar1} we obtain:
\begin{gather}
    \begin{split}
    \psi_{1}(x) &= \xi_{1,+}e^{i\int_{z_p}^{x_0}q_+dx}
    \Big(e^{i\int_{x_0}^xq_+dx}
    \\
    & +\frac{i\sqrt{\pi}\mu}{2\ve^{3/2}\sqrt{a}}
    e^{-i\int_{z_p}^{x_0}q_+dx} e^{+i\int_{z_p}^{x_0}q_-dx}e^{+i\int_{x_0}^{x}q_-dx}\Big)
\end{split}
\end{gather}
where $x_0$ is an arbitrary point on the real axis. This way we obtain the reflection amplitude
\begin{gather}
r=\sqrt{\frac{\pi}{a}}\frac{\mu}{2\ve^{3/2}}\exp\left[i\int_{z_p}^{x_0}(q_--q_+)\,dx\right].
\end{gather}
And for the reflection coefficient, $R=|r|^2$, we finally have the following expression:
\begin{gather}
\label{eq:R}
    R=\frac{\pi\hbar v_F}{|a|}\frac{\mu^2}{4\ve^3}
    \exp\!\!\left[-\frac{4}{\hbar v_F}{\rm Im}\int\limits_{x_0}^{z_p}
    \frac{\sqrt{\ve^2-\mu^2+\ve^2\vf^2(z)}}{1+\vf^2(z)}dz\right]
\end{gather}
Here we have restored $\hbar$ and Fermi velocity.
Result~\eqref{eq:R} complements result~\eqref{reflection2} obtained for a regular deformation function $\vf(z)$. 
We see that the two relations share the same semiclassical exponent, as should be expected from the semiclassical analysis. The difference between different potential classes is pronounced in the pre-exponential factor. And here is where the first interesting feature of the problem comes on stage. 

Comparing reflection coefficients ~\eqref{reflection2} and ~\eqref{eq:R}, we see that, unlike the former, the latter
explicitly respects the TR symmetry of the TI. Namely, the reflection coefficient vanishes in the case of vanishing magnetic field $\mu\rightarrow0$, when the TR symmetry of quasiparticle excitations of TI is restored. 

\section{Born approximation}
\label{born}
The aim of this part of the paper is to resolve the paradox of non-vanishing at zero magnetic field reflection coefficient, outlined in the previous section. 
As a starting point, we need to analyze the scattering problem in the weak magnetic field limit $\mu\ll\ve$, restricting ourselves to the first Born approximation.

TR symmetry of the problem gives us a nice present here. Remarkably, we managed to find an exact solution of the Dirac equation~\eqref{dirac:main} in the absence of the magnetic field $\mu = 0$ for any deformation potential. Naturally, due to TR symmetry, the exact solution is reflectionless.  
Now we are going to see, how even the slightest magnetic field affects the analytical structure of the solution and leads to non-zero reflection in the problem. 
\subsubsection{Exact solution}
Let us rewrite the initial Hamiltonian in the absence of magnetic field:
\begin{gather}
\label{ham-free}
H=v_F\sigma_y\hat p+\frac{\sigma_z}{2}(\vf\hat p+\hat p\vf)
\end{gather}
It happens one can contrive a unitary transformation 
\begin{align}
    \psi(x) &= \hat{U}(x)\tilde{\psi}(x),\\
    \hat{U}(x)&=\exp [i\theta(x)\sigma_x],\\
    \tan2\theta(x)&=\vf^{-1}(x),\label{theta}
\end{align}
turning  Hamiltonian~\eqref{ham-free} to much simpler form (see Appendix~\ref{app:trans}):
\begin{gather}
\label{ham_exact}
\tilde{H}=\frac12(v\hat p+\hat p v)\sigma_z,
\end{gather}
where $v(x)=v_F\sqrt{\vf^2(x)+1}$.
Hamiltonian~\eqref{ham_exact} has the following exact eigenfunctions (see the derivation in~\ref{app:eigen}):
\begin{align}
\label{sol_exact}
\underset{\rightarrow}{\psi_\ve}(x)&=\frac{e^{i\ve\tau(x)} }{\sqrt{v(x)}}\begin{pmatrix}1\\0\end{pmatrix}, \quad
\underset{\leftarrow}{\psi_\ve}(x)=\frac{e^{-i\ve\tau(x)}}{\sqrt{v(x)}}\begin{pmatrix}0\\1\end{pmatrix},
\\
\label{tau0}
\tau(x)&=\int\limits_{0}^{x}\frac{dx'}{v(x')}
\equiv \int\limits_{0}^{x}\frac{dx'}{\sqrt{\vf^2(x^\prime)+1}}.
\end{align}
And one clearly sees that the forward moving exact solution in~\eqref{sol_exact} remains such  in the entire real axis and we have the reflectionless situation expected from the TR symmetry of the system.
\subsubsection{Perturbation theory in $\mu$.}
To build the perturbation theory, we need   
the Green's function for the transformed Hamiltonian~\eqref{ham_exact} (see Appendix~\ref{green}):
\begin{gather}
\label{green_funa}
G(\epsilon;x,x')=
-\frac{i}{2}
(1+\text{sign}[\tau(x)-\tau(x')]\sigma_z)
\frac{e^{i\epsilon|\tau(x)-\tau(x')|}}{\sqrt{v(x)v(x')}},
\end{gather}
where ${\rm sign}\,(x)$ is a sign function.
Then we consider the perturbation created by magnetic field; in the initial basis it is $V=\mu\sigma_z$. Under the unitary transformation $\hat U$ it becomes:
\begin{gather}
\label{pot_transf}
\tilde{V}(x)=\frac{\mu}{\vf^2(x)+1}\left[\vf(x)\sigma_z-\sigma_y\right]
\end{gather}
Then, the reflected wave is given by the perturbation theory:
\begin{gather}
\label{perturb}
    \psi_{\rm ref}(x) = -\int\limits_{-\infty}^\infty G(\epsilon;x,x^\prime)\tilde{V}(x^\prime)\underset{\rightarrow}{\psi_\ve}(x^\prime)\,dx^\prime
\end{gather}

Plugging  the transformed scattering potential~\eqref{pot_transf}, the Green's function~\eqref{green_funa} into ~\eqref{perturb},
we obtain (after some simple algebra) the reflected wave in the first order perturbation theory:
\begin{gather}
\label{born_fin}
\psi_{\text{ref}}= r\underset{\leftarrow}{\psi_\ve}(x),\quad r=\mu\int\limits_{-\infty}^{\infty}\frac{e^{2i\ve\tau(x')}}{1+\vf^2(x')}dx'
\end{gather}
where $r$ is the final reflection amplitude in Born approximation. 
A shrewd reader is going to immediately notice that the integral defining $r$ is divergent. 
As argued in Appendix~\ref{app:eigen}, one should understand this integral as a taken along the inclined directions $-\infty\rightarrow \infty e^{i\pi-\delta}$ and $\infty\rightarrow\infty e^{i\delta}$ where $\delta$ is an arbitrarily small positive angle.

\section{Matching Born approximation with semiclassics}
\label{match}
Part~\eqref{born} was dedicated to the derivation of the reflection amplitude in the first Born approximation. However,
Born reflection amplitude~\eqref{born_fin} ought to be valid even in the semiclassical limit provided the semiclassical condition ~\eqref{semi_cond} is satisfied.

As it is easy to see, condition~\eqref{semi_cond} implies the evaluation of Born amplitude~\eqref{born_fin} with the steepest descent technique. Therefore, we expect the reflection given by the steepest descent integration in the Born approximation and semiclassical answers should match. Following the general structure of our semiclassical exposition, we need to address two substantially different cases: the case of regular profile potential $\vf(z)$  (Sec.~\ref{reg1}) and the potential with a pole  (Sec.~\ref{pole}).

\subsection{Regular potential}
\subsubsection{Semiclassical limit of the Born approximation}
As in Sec.~\ref{reg1}, function $\vf(z)$ is assumed to be an entire function in the complex plane. Therefore, function $\tau(z)$ (Eq.~\ref{tau0}) has no singularities apart from the square-root-type branch points  where function $\vf^2+1$ has roots (which are assumed to be of the first order in a general situation).

The steepest descent technique implies the deformation of the contour in integral~\eqref{born_fin} along the steepest descent paths of the real part of the exponential function: ${\rm Re}\,i\tau(z)$.
Therefore, we need to know the structure of the stationary curves of the function $\tau(z)$, i.e., the level lines of ${\rm Re}[\tau (z)]$.  The detailed analysis is a little involved but reveals a sight to behold. The resulting contour circumvents each of the branch points of  $\tau(z)$ twice round infinitesimal circles. The global placement of the contour along the steepest descent paths spans infinite amount of Riemann sheets of $\tau(z)$. The details are presented in Appendix~\ref{app:analyt}.
Here, we present the summary of our observations. For exposition's clarity, we start with addressing the case of a single branch point $z_0$ (in the upper complex half-plane). 

\begin{figure}[t!]
	\centering	\includegraphics[width=0.85\columnwidth]{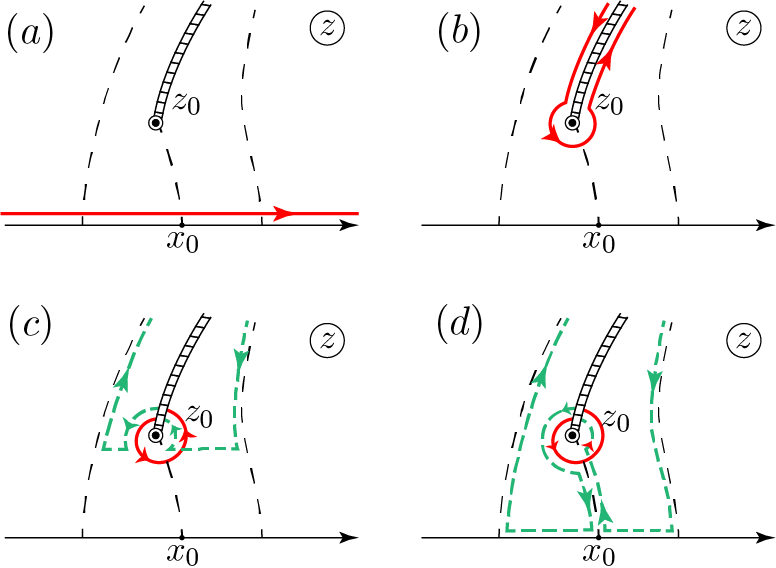}
	\caption{(a-d) The steepest descent paths 
 (black dashed lines) and the deformation of the contour along the steepest descent paths  of function ${\rm Re}i\tau(z)$. Color code of the contour means its placement on a different Riemann sheet of $\tau(z)$.}
\label{fig:deform_init}
\end{figure}

\begin{figure}[t!]
	\centering	\includegraphics[width=0.85\columnwidth]{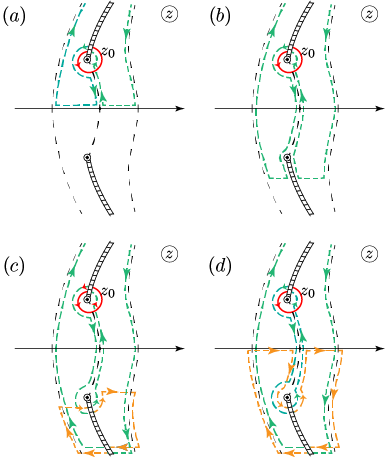}
	\caption{(a-d) The iterative deformation of the contour along the steepest descent paths of function $\tau(z)$. The steepest descent paths are denoted as black dashed lines. The  color code of the contour denote its belonging to a respective Riemann sheet. One notices that at the end of the procedure the deformation again has horizontal segments. But they are positioned on the Riemann sheet with even lower values of ${\rm Re}\,i\tau(z)$ than the segments in the previous figure~\ref{fig:deform_init}. Therefore, repeating this step iteratively, the contribution from the horizontal paths becomes exponentially smaller and smaller until it becomes $e^{-\infty} = 0.$}
\label{fig:deform2_inf}
\end{figure}

\begin{figure}[t!]
	\centering	\includegraphics[width=0.85\columnwidth]{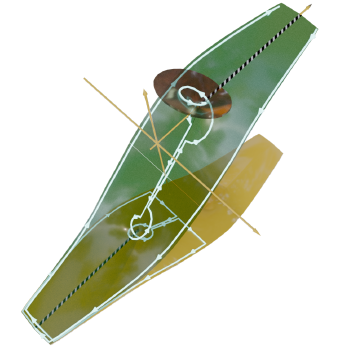}
	\caption{The 3D visualization of the ${\rm Re}i\tau(z)$ with the profile function $\vf(z)=z e^{-z^2/2}$. To make things more readable, we draw a small circular region of the upper Riemann sheet (the red one), which supports the placement of the contour (the infinitesimal circle)}
\label{fig:deform2_inf3D}
\end{figure}

The branch cut should be drawn in  the upward direction ${\rm Re}[\vf(z)^2]\rightarrow-\infty$. 
(See Fig.~\ref{fig:deform_init}(a)). There is always the steepest descent path starting at the real axis and ending at branch point $z_0$. There are also two steepest descent paths to the left and to the right, stretching all the way to $+i\infty$. The simplified deformation of the contour spans two Riemann sheets of function $\tau(z)$ and is presented in Fig.~\ref{fig:deform_init}(d). 
Despite the fact that this placement leads to the correct asymptotic of the integral~\eqref{born_fin} it is slightly inaccurate.  It contains two horizontal segments on the real axis, which corresponds to the level lines of ${\rm Re}\,[i\tau(z)]$ rather than its steepest descent lines (i.e. the lines of maximal oscillation of $\exp[2i\ve\tau(z)]$). Fortunately, these segments are positioned on the Riemann sheets of function $\tau(z)$ with smaller values of ${\rm Re}\,i\tau(z)$ leading to an exponentially small correction to the estimate of the integral~\eqref{born_fin}. The inquisitive reader may ask what would be the correct placement of the contour going exclusively along the steepest descent paths? The answer is: the correct contour placement spans an infinite amount of Riemann sheets. It is easy to understand the position of the contour as an iterative procedure (see the explanation and 3D visualization in Figs.~\ref{fig:deform2_inf}, ~\ref{fig:deform2_inf3D}).

\begin{figure*}[t!]
	\centering
\includegraphics[width=2\columnwidth]{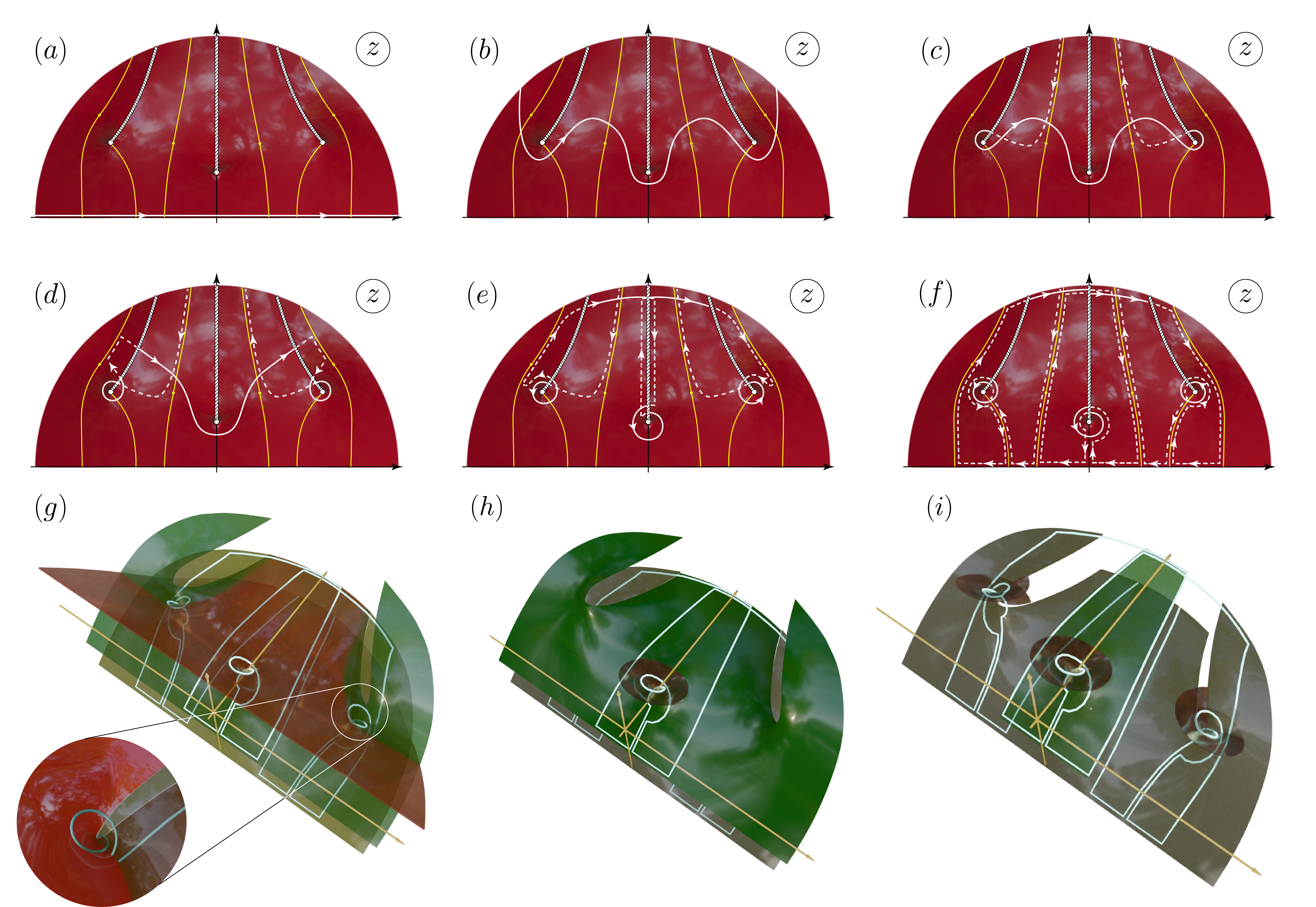}
	\caption{(a-f) The deformation of the contour for the function $\tau(z)$ for the profile deformation $\vf(z) = ze^{-z^2/2}$. The final placement of the contour (f) corresponds to circumventing each branch point twice. (g) the 3D visualization of the three Riemann sheets of the surface of the function $\tau(z)$ (the real part of $\tau(z)$). Each sheet is drawn semitransparent to reveal the whole placement of the contour. (h) The green surface is presented with solid color. To avoid ambiguity with the placement of the contour, the  part of the upper sheet supporting the contour is drawn as a circular inset at the central branch point $z_0$. It is the Riemann sheet connected to the upper sheet (the red one) via the central branch cut (along the imaginary axis). (i) The lowest Riemann sheet is depicted. It is connected to the upper sheet via side branch cuts. The inset from the Green sheet supporting the contour as well as circular insets from the upper sheet are also present.}
\label{fig:stokes2}
\end{figure*}

What we discussed so far was the unrealistic case of the of $\tau(z)$ having a single branch point. Simple arguments (Appendix \ref{app:analyt}) show that $\tau(z)$ always has an infinite number of branch points. Therefore, the deformation of the contour for a typical $\tau(z)$ is presented in Fig.~\ref{fig:stokes2}. 
The principal contribution comes from the closest to the real axis branch point. Only three branch points happen to be inside the captured area of the Riemann surface of $\tau(z)$ in Fig.~\ref{fig:stokes2}. However, we hope the reader grasps the general idea of the contour deformation from the illustration.

As we see from Fig.~\ref{fig:stokes2} the integral along the steepest descent contours sprawling from the branch points cancel each other. The contributions from other steepest descent lines are exponentially suppressed, since they are positioned on the lower Riemann sheets. As a result, the leading contribution to the integral is given by the sum of the residues of the integrand in~\eqref{born_fin} multiplied by $4\pi i$ (double circumvention).  
\begin{gather}
\begin{split}
    r &\underset{\ve a\rightarrow\infty}\longrightarrow 4\pi i\mu\sum\limits_{\rm all\ branch\ points}\underset{z_n}{\rm res} \frac{e^{2i\ve\tau(z)}}{\vf^2(z)+1}\\
    &\approx 2\pi a\mu e^{2i\ve\tau(z_0)}.
\end{split}
\end{gather}
where $a$ is defined as the characteristic potential length via the Taylor expansion
\begin{gather}
\label{pole2} 
    \vf(z) = i+(z-z_0)/a.
\end{gather}
As a result, the reflection coefficient in the Born approximation in the semiclassical limit:
\begin{gather}
\label{scattering_perturb}
    R \equiv |r|^2 = \frac{4\pi^2\mu^2|a|^2}{v_F^2\hbar^2}\exp\left(-\frac{4\ve}{\hbar v_F}{\rm Im\int\limits_0^{z_0}}\frac{dz}{\sqrt{\vf^2(z)+1}}\right). 
\end{gather}

\subsubsection{Born Approximation from the semiclassical analysis}
Comparing formula for the reflection coefficient presented by P-Kh analysis~\eqref{reflection2} and the answer~\eqref{scattering_perturb} given by the semiclassical limit of perturbation theory, 
one immediately sees there is no way semiclassical result~\eqref{reflection2} can match the perturbative expression~\eqref{scattering_perturb}. This can be vividly seen from the fact that, unlike~\eqref{scattering_perturb}, expression~\eqref{reflection2} has nonzero limit at $\mu\rightarrow0$. 

The problem is, the type of semiclassical analysis undertaken in Sec.~\ref{subsec:branch} breaks down when $\mu \rightarrow 0$. The reason for this is that at $\mu\ll\ve$ the pole of the semiclassical momentum~\eqref{mom_semi} at $\vf = \pm i$ approaches the branch points of $p$: $\vf_\pm$ (see Eq.~\ref{branch1}).

This means, the exact solution near the turning point~\eqref{exact1} can't be continued to the asymptotic region~\eqref{region_semi} because the Taylor expansion of function $\vf(z)$ near the turning point stops being valid due to the presence of the pole $z_0: \vf = i$. The necessary estimate is easy to make. When $\mu\ll\ve$ the distance $z_0-z_\pm\ll a$, Therefore:
\begin{align}
    \vf(z_0)-\vf(z_\pm)\sim\frac{\mu^2}{\ve^2}\sim\frac{z_0-z_\pm}{a} \ \Rightarrow\\
    \label{dist}
    |z_0-z_\pm|\sim \frac{\mu^2}{\ve^2}a
\end{align}
The crucial asymptotic expansion~\eqref{exact1} becomes invalid if  the distance $s\sim 1$ in\eqref{exact1} (equivalently, $|z-z_\pm|\sim a^{1/3}/\ve^{-2/3}$ is of the same order as distance~\eqref{dist}. That means
\begin{gather}
\label{mu_small}
    \frac{\mu}{\ve}\sim\frac{1}{(a\ve)^{1/3}}.
\end{gather}
Therefore, at $\mu\lesssim \ve^{2/3}/a^{1/3}$ the regular P-Kh method breaks down.
The possible modification of the method, when the branch point and the pole start to coalesce, is also outlined (for the Schrödinger equation) in~\cite{pokr1961}.
First, we point the reader's attention to the structure of the perturbative result~\eqref{scattering_perturb}.
It shows that the main contribution to the scattering amplitude is given by the semiclassical action $S = \int_{x_0}^{z_0}\pi(x)\,dx$ (see expression for $\pi$ in Eq.\ref{semi_cond} at $\mu = 0$), hinting that the over barrier scattering happens at branch point $\vf = i$ (i.e. $z_0$). In fact, as we are going to see, the Stokes lines still sprawl out of the branch points $z_\pm$ but get strongly distorted by the presence of the pole $z_0$.

\begin{figure}[H]
	\centering	\includegraphics[width=1\columnwidth]{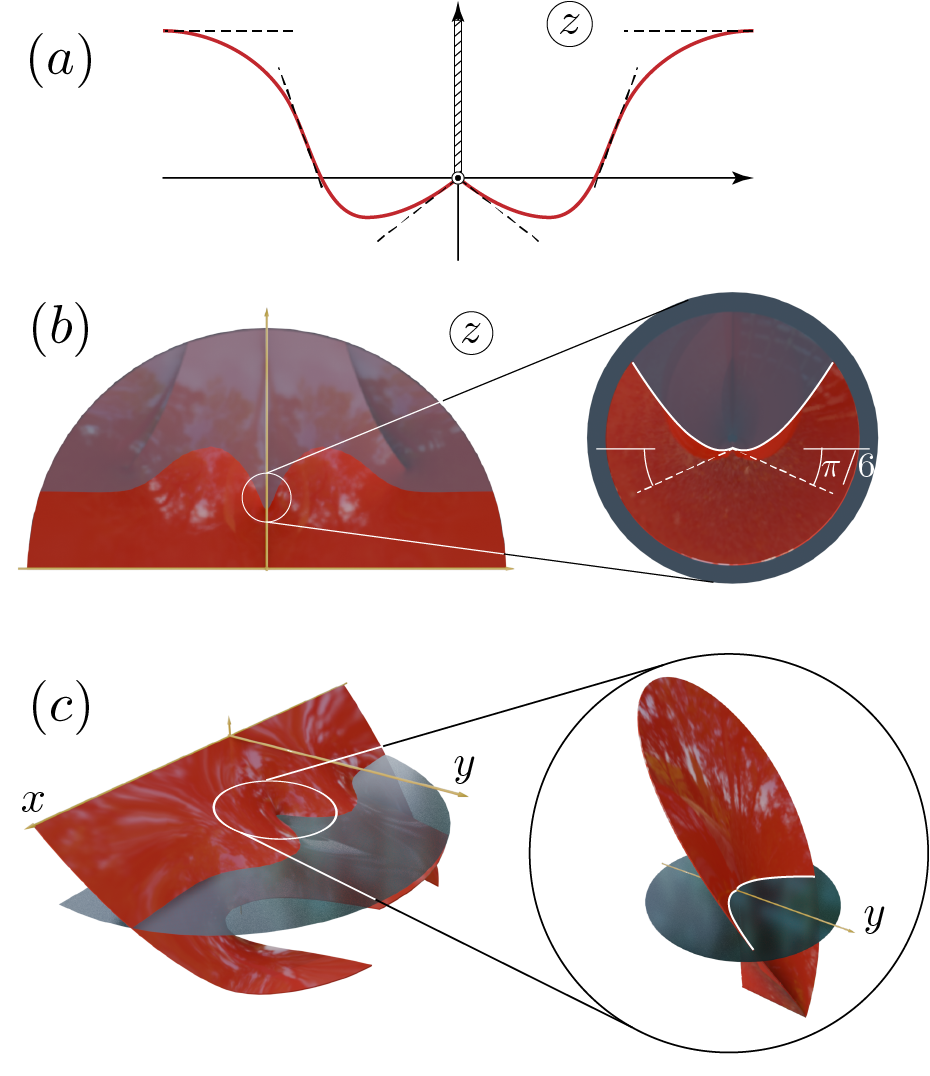}
	\caption{(a) The structure of the anti-Stokes line from the general considerations. (b-c) The  anti-Stokes lines are depicted as level lines of function $i{\rm Im}\int_0^\zeta\pi(t)dt$ with profile function $\vf(z)=z e^{-z^2/2}$ for the $\mu/\ve = 0.15$.}
\label{fig:principal1}
\end{figure}

\paragraph{Anti-Stokes lines at the branch point of $p$ at $\mu\ll\ve$}
We implement step (ii) of P-Kh method. As before, we choose point $z_+$. 
However, this time one should use Laurent expansion of the semiclassical momentum~\eqref{semi3} near the pole $z_0$ rather than Taylor expansion near the branch point $z_+$. Plugging in expansion~\eqref{pole2}, we obtain:
\begin{gather}
   \label{semi-expansion2}
    \pi(\zeta) = \frac{a\ve}{2i\zeta}\sqrt{\frac{2i\zeta}{a}-\frac{\mu^2}{\ve^2}}+...\rightarrow e^{-i\pi/4}\ve\sqrt{\frac{a}{2\zeta}}.
\end{gather}
The last transition in~\eqref{semi-expansion2} is made under assumption $|\zeta/a|\gg (\mu^2/\ve^2)$.
The semiclassical action becomes:
\begin{gather}
    \int\limits_0^\zeta\pi(t)\,dt = e^{-i\pi/4}\ve\sqrt{2a\zeta}.
\end{gather}
Therefore, the semiclassical approximation in this case corresponds to the limit:
\begin{gather}
     \label{semi2}
    |\zeta|\gg\frac{1}{|a|\ve^2},
\end{gather}
and the condition for the anti-Stokes line becomes now
\begin{gather}
\label{anti-Stokes2a}
    {\rm Re}\,i \int\limits_0^\zeta\pi(t)\,dt = 0\ \Rightarrow\ \arg (a\zeta) =\frac{\pi}{2}+2\pi n,\ n\in\mathbb{Z}. 
\end{gather}
Eq.~\eqref{anti-Stokes2a} together with ~\eqref{anti-Stokes1} presents a rather peculiar pattern of anti-Stokes contours.
In the close vicinity of branch point $z_+$ (where the Taylor expansion of semiclassical momentum~\eqref{exp1} is valid)  the anti-Stokes lines form a triplet sprawling out at $2\pi/3$ angles from each other. On the other hand, Eq.~\ref{anti-Stokes2a} tells us that they merge into a doublet, passing along the banks of the branch cut defining a single valued branch of $\pi(\zeta)$. Finally, we expect the anti-Stokes lines shape into a doublet consisting of two horizontal lines when far enough from $z_0,\ z_\pm$.      
These general considerations are illustrated in Fig.~\ref{fig:principal1}(a) for the case when the expansion parameter $a$ in~\eqref{branch2} and in~\eqref{pole2} is real (for concreteness).  

For concreteness from now on, we will assume that parameter $a$ in expansion~\eqref{pole2} is real. 
For the general situation of non-zero $\arg a$, the only difference would be the inclination of all the graphics by the necessary angle.

The pattern looks rather unusual (and, to some extent, even improbable). To illustrate its correctness, we present the drawing of exact anti-Stokes lines (as a level line of the relief of ${\rm Re} i \int_{z_+}^z\pi(t)\,dt$  in Fig.~\ref{fig:principal1}(b-c) for a specific potential function $\vf(z)=z \exp(-z^2/2)$.
We are interested in asymptotics of solutions on anti-Stokes lines. 
However, as one looks at the structure of exact anti-Stokes lines presented in Fig.\ref{fig:principal1} and its approximate direction given by analytical relation ~\eqref{anti-Stokes2a}, one inevitably notices an obvious discrepancy (due to non-zero $\mu$ discarded in Eq.~\ref{anti-Stokes2a}). How then can one be sure in the validity of our treatment if the found solution would correspond to the anti-Stokes line, which only approximately resembles the exact one? 

To address this concern, a short comment on asymptotic analysis seems appropriate and is presented in Appendix~\ref{app:asymp}.
The short answer is, we don't necessarily need to consider the solution \textit{exactly} on the anti-Stokes line, since the asymptotics of the solution remains valid also in the vicinity of the anti-Stokes line due to the principle of analytic continuation (as long as this vicinity doesn't touch the Stokes-line). 

Semiclassical solutions~\eqref{hbar1} now are easy to obtain taking the starting point $z=z_0-ai\mu^2/(2\ve^2)$ (see Appendix~\ref{app:semi_mu}). 
\begin{align}
\label{sem_reg1}
    \psi_{+,>}&=\sqrt{\ve}\left(\frac{\mu}{\ve}\right)^{ia\mu}
    \left(\frac{a}{2y}\right)^{\frac{ia\mu}{2}+\frac{1}{2}}e^{i\ve\sqrt{2ay}-i\frac{\pi}{4}+\frac{\pi\mu a}{2}}\\
\label{sem_reg2}    
       \psi_{-,<}&=\sqrt{\ve}\left(\frac{\mu}{\ve}\right)^{ia\mu}
    \left(\frac{a}{2y}\right)^{\frac{ia\mu}{2}+\frac{1}{2}}e^{i\ve\sqrt{2ay}+\frac{i\pi}{4}-\frac{\pi\mu a}{2}}\\
\label{sem_reg3}
        \psi_{+,<}&=\sqrt{\ve}\left(\frac{\mu}{\ve}\right)^{ia\mu}
    \left(\frac{a}{2y}\right)^{\frac{ia\mu}{2}+\frac{1}{2}}e^{-i\ve\sqrt{2ay}+\frac{3i\pi}{4}-\frac{\pi\mu a}{2}}.
\end{align}
Here, notation index $\pm$ correspond to~\eqref{hbar1}.

\paragraph{Exact solution near the branch point $p$ at $\mu\ll\ve$}
Expanding the Dirac equation~\eqref{main} near $z_0$ we obtain:
\begin{gather}
\label{bessel1}
    2i\zeta\psi_1^{\prime\prime}+(3i-2a\mu)\psi_1^\prime+\ve^2a\psi_1=0.
\end{gather}
Eq.~\ref{bessel1} is the Bessel type equation. In principle, one may look up its analytic properties and the structure of its asymptotics at different arguments. However, to study the analytical continuation of its solutions
it is instructive to present its exact solution via Laplace method (see Appendix)

\begin{gather}
\label{exact2}
    \psi_1(\zeta) =-i\ve\sqrt{\frac{\ve a}{2\pi}}\left(\frac{\mu}{\ve}\right)^{ia\mu}e^{\frac{\pi a \mu}{2}} \int\limits_C e^{\zeta\ve s+\frac{\ve a i}{2s}}s^{ia\mu-\frac{1}{2}}\,ds.
\end{gather}
Here, the constant in front of the integral is tuned in such a way that the asymptotics of $\psi(iy)$ in~\eqref{exact2} on the right anti-Stokes line matches the respective semiclassical solution~\eqref{sem_reg1}. 

\begin{figure}[H]
	\centering	\includegraphics[width=1\columnwidth]{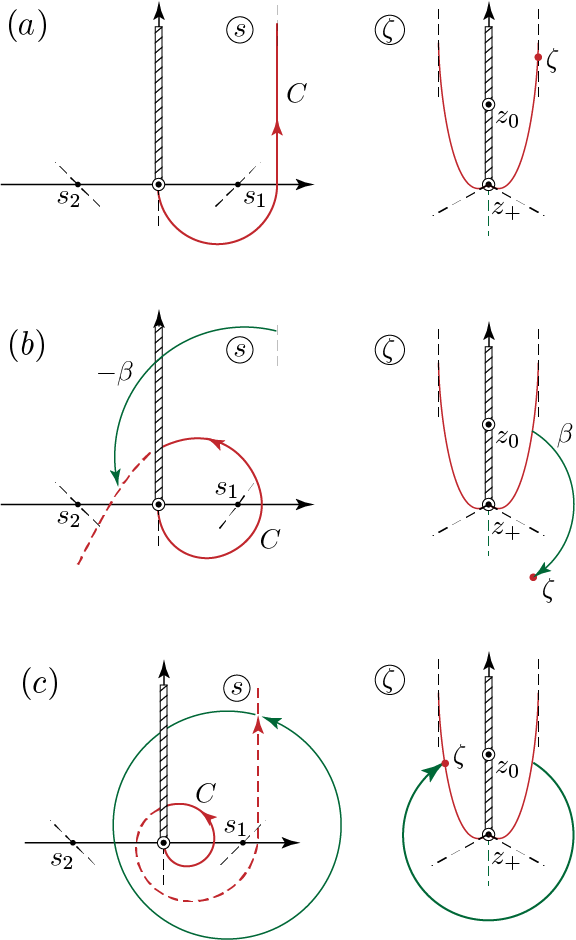}
	\caption{(a) 
 Right: anti-Stokes lines near the point $z_+$. Point $\zeta$ on the right anti-Stokes line is such that the wave function at this point represents the outgoing wave and should match ~\eqref{sem_reg1}. The dashed lines are the approximate directions of anti-Stokes lines.
 Left: the initial placement of the contour $C$ for solution~\eqref{exact2}. $s_1$ and $s_2$ are the saddle points of function $f(s,\zeta)$ in ~\eqref{func}. The dashed lines are the steepest descent lines of $f(s,\zeta)$. 
 (b) Right: Argument $\zeta$ of the wave function is in transition from the right anti-Stokes line to the left one. It is rotated by angle $\beta$. Left: The red dashed line denotes the placement of the contour on a different Riemann sheet of multivalued integrand in ~\eqref{exact2}. The steepest descent direction of $f(s,\zeta)$ at $s\rightarrow\infty$ is now inclined by angle $-\beta$. (c) Right: Argument $\zeta$ is on the left anti-Stokes line (is rotated by $2\pi$). Left: The contour $C$ is deformed into a spiral.}
\label{fig:deform3}
\end{figure}

The contour which gives the correct asymptotics at the right anti-Stokes line $\zeta = iy$ is presented in Fig.~\ref{fig:deform3}(a). The placement of the contour is dictated by the possibility to draw the steepest descent line of the exponent function in~\eqref{exact2}:
\begin{gather}
\label{func}
    f(s, \zeta) = \zeta s +\frac{ai}{2s}
\end{gather}
through the saddle $s_1 = \sqrt{a/2y}$ ($\zeta = iy$) of function $f(s, \zeta)$. This way, exact solution~\eqref{exact2} gives the correct main exponential (see the semiclassical expression~\eqref{sem_reg1}).

\begin{figure}[t]
	\centering	\includegraphics[width=1\columnwidth]{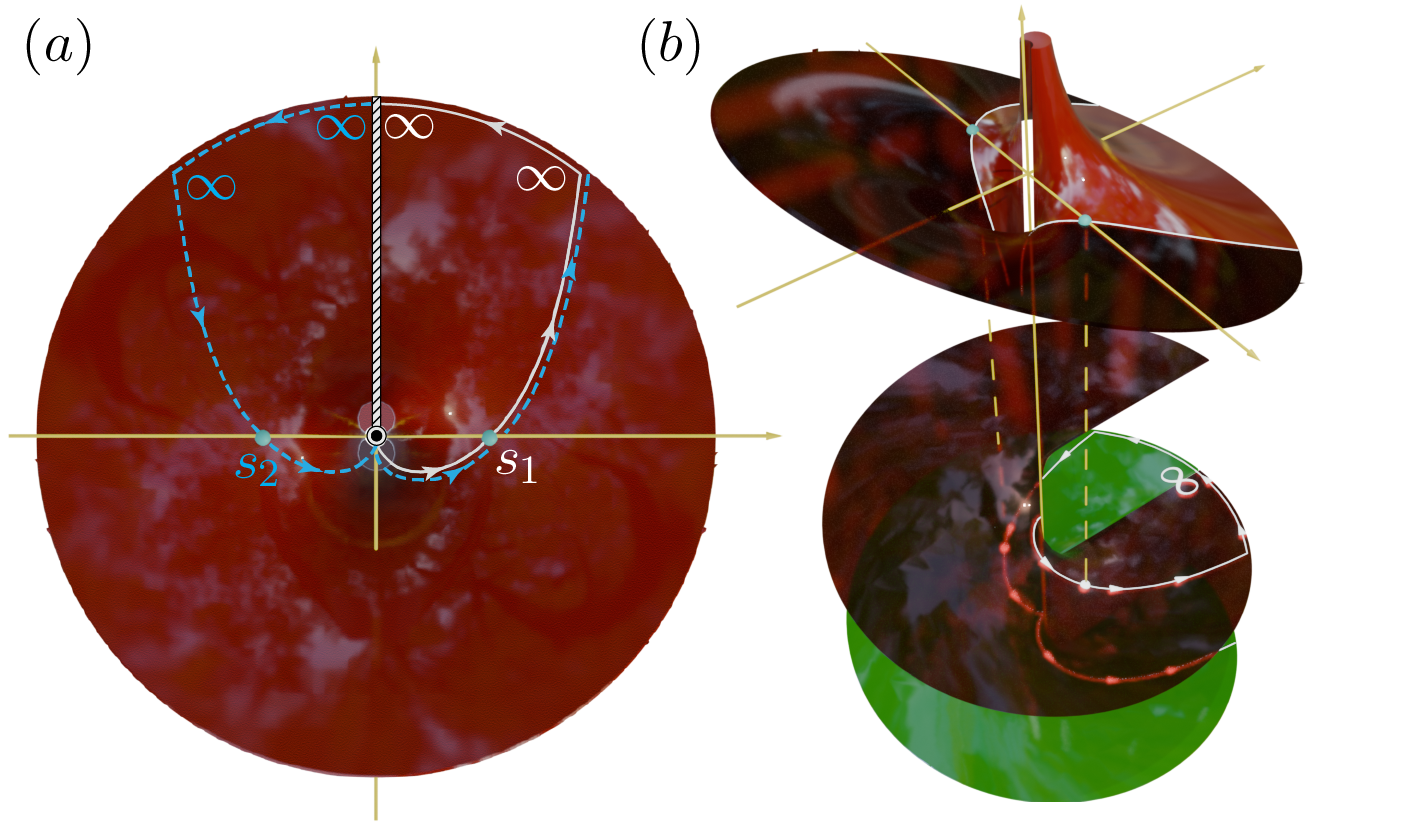}
	\caption{(a) The steepest descent contour, giving the asymptotics of the exact solution~\eqref{exact2} on the left anti-Stokes line. The cyan color means the placement of the contour on a different Riemann sheet of the multivalued function $s^{ia\mu-1/2}$ from the integrand~\eqref{exact2}. (b) Up: The placement of the contour on the relief of function ${\rm Re}f(s, \zeta)$.
     Down: The placement of the contour on the Riemann surface of the function $s^{ia\mu-1/2}$. The red sheet corresponds to the regular branch $s^{ia\mu-1/2} \equiv \exp((ia\mu-1/2)\ln|s|)$ for $s>0$ while the green sheet corresponds to $s^{ia\mu-1/2} \equiv \exp([ia\mu-1/2][\ln|s|+2\pi i])$ for $s>0$.}
\label{fig:principal3}
\end{figure}

After one rotates complex number $\zeta$ from the right to the left anti-Stokes line in the clockwise direction, the contour $C$ defining an exact solution ~\eqref{exact2} has to rotate by the same angle in the counterclockwise direction to guarantee the convergence of the integral (see Fig.~\ref{fig:deform3}(b)). At the same time, the starting direction of the contour ($s=0$) has to be kept downwards (again, for the integral to converge).
As a result, contour $C$ is swirled into a spiral (see Fig.~\ref{fig:deform3}(c)) after $\zeta$ arrives at the left anti-Stokes line.
As argued in Appendix~\ref{app:asymp}, for the procedure of asymptotics computation to be flawless, the integration contour $C$ has to be drawn along the global steepest descent paths of $f(s,\zeta)$. The respective deformation is presented in Fig.~\ref{fig:principal3}(a-b). 
As we see, the right saddle $s_1$ contributes twice, while integration contour $C$ passes it on different Riemann sheets of the integrand. The contribution from this saddle, therefore, gets multiplied by factor $1+e^{2\pi i(ia\mu-1/2)} \equiv e^{-\pi\mu a}2\sinh\pi\mu a$. 
As a result (see Appendix~\ref{app:fin} for simple details), the asymptotics of~\eqref{exact2} on the left anti-Stokes line can be expressed via semiclassical functions~\eqref{sem_reg2}, ~\eqref{sem_reg3} at the same line as follows. 
\begin{gather}
\label{victory1}
    \psi(iy)\Big|_{\rm left} =\psi_{+,<}(iy)-2i\sinh\pi\mu a \psi_{-,<}(iy)
\end{gather}
Eq.~\ref{victory1} gives the ability to match exact solution~\eqref{exact2} in the vicinity of the branch point with the semiclassical solutions in the entire complex upper half-plane, including the real axis. 

\paragraph{Matching semiclassical and exact solutions}
Proceeding along step (v) of the P-Kh method we obtain:
\begin{gather}
\label{victory2}
\begin{split}
    \psi(x)&\underset{x\rightarrow-\infty}{=}\xi_{1+}\exp\left(i\int\limits_{z_+}^x q_+dz\right)\\
    &-2i\sinh\pi\mu a \xi_{1-}\exp\left(i\int\limits_{z_+}^x q_-dz\right)
\end{split}
\end{gather}
Finally, restoring the Planck's constant and Fermi velocity and changing $a\rightarrow |a|$, we obtain for the reflection coefficient:
\begin{gather}
\label{victory3}
    R\!=\!4\sinh^2\frac{\pi\mu |a|}{\hbar v_F}\exp\!\!\left(\!-\frac{4}{\hbar v_F}{\rm Im}\!\!\int\limits_{x_0}^{z_+}\frac{\sqrt{\ve^2(\vf^2+1)}-\mu^2}{\vf^2+1}\,dz\right)
\end{gather}
Reflection coefficient~\eqref{victory3} represents the generalization of result~\eqref{reflection2} on the case $\mu a\sim1$.
Result~\eqref{victory3} beautifully matches the Born approximation result~\eqref{scattering_perturb} in the limit $\mu a\rightarrow 0$. Eq.~\eqref{victory3} also resolves the paradox of non-vanishing at $\mu\rightarrow0$ reflection coefficient~\eqref{reflection2}.


\subsection{Potential with a pole}

Now we implement the same program for much simpler case of potential with a pole.
The reflection amplitude  obtained in~\eqref{born_fin} should be matched with the one obtained from the semiclassical approximation~\eqref{eq:R} in the limit $\ve a/(\hbar v_{\rm F})\gg1$. As before, this corresponds to the saddle point evaluation of the integral entering~\eqref{born_fin}:
\begin{gather}
\int\limits_{-\infty}^{\infty}\frac{e^{2i\ve\tau(x')}}{1+\vf^2(x')}dx'
\approx
e^{i\pi/4}\sqrt{\frac{\pi}{\ve a}}\frac{e^{2i\ve\tau(z_p)}}{2\ve}
\end{gather}
where $z_p$: $\vf(z_p)=\infty$ is the saddle point of the function $\tau(z)$ (which also is the turning point for the semiclassical momentum). This way we get for the reflection coefficient:
\begin{gather}
R=\frac{\pi\mu^2}{4|a|\ve^3}\exp \Big[-4\ve\;\text{Im}\int\limits_{0}^{z_0}\frac{dx}{\sqrt{1+\vf^2(x)}}\Big]
\end{gather}
This result coincides with the semiclassical reflection coefficient (\ref{eq:R}) in the limit $\mu\ll\ve$. This match provides an additional convincing proof of the correctness of the semiclassical result~\eqref{eq:R}

Reflection coefficients ~\eqref{reflection2}  and ~\eqref{eq:R} obtained in the semiclassical approximation for two wide classes of general type deformations, complemented by refined result~\eqref{victory3} conclude our study of the scattering phenomena in the problem. They are the main results of this paper.

\section{Discussion}
\label{discussion}
To conclude, we studied analytically the scattering of the quasiparticles on edge imperfections  of 2D TI in the uniform magnetic field. We used two complementing each other approaches: Pokrovsky-Khalatnikov method and perturbation theory in magnetic field. We obtained the reflection coefficients for two most important wide classes of deformation potentials and made sure they match in the shared domain of validity of both treatments. The study reveals the nontrivial interconnection between TR symmetry and the analytical properties of the reflection amplitude. 

Our results allow for a direct experimental check.
The perturbation theory results are obviously valid for sufficiently small external magnetic field. The semiclassical parameter $\lambdabar/a=\hbar v_F/(\ve a)$ 
is easy to estimate from typical experimental data. For 2D TI formed in gated HgTe quantum
well, the Rahsba splitting parameter $\alpha \sim 10$ eV\AA, \cite{schultz1996rashba}, 
the Fermi velocity $v_F\approx 2$ eV\AA,~\cite{FermiVel}. We see that Rashba parameter  $\alpha$ is approximately of the same order as Fermi velocity $\alpha\sim v_R$. Therefore, for the typical experiment, the $1\mu$ size edge defect exceeds by far the quasiparticle wave length $\lambda\sim 100$\AA,   \cite{FermiLevel} which justifies the use of semiclassical approximation.

\section*{Acknowledgments}
The authors acknowledge support by the Federal Academic Leadership Program "Priority 2030" , NUST MISIS Grant K2-2022-025, the Foundation for the Advancement of Theoretical Physics and Mathematics ”Basis” for grant $\#$ 22-1-1-24-1, PDG acknowledges State Assignment $\#$ FFWR-2024-0015.

\appendix

\section{Semiclassical equation}
\subsection{Derivation of the main semiclassical exponential and pre-exponential term}
\label{app:semi-sol}
We plug in \eqref{psiseries} into the main differential equation~\eqref{main} and perform the formal expansion in powers of $\hbar$. Neglecting the terms of order $\hbar$ (the zeroth order semiclassical expansion) results in the following quadratic equation on $S_0'(x)$:
\begin{gather}
\label{zero-order}
4(\ve+\mu)[(1+\vf^2)(S_0')^2+2\mu\vf S_0']=4(\ve+\mu)(\ve^2-\mu^2)
\end{gather}
which solution is written in the main text (\ref{mom_semi}). Next, keeping the next terms (of the order of $\hbar$) we get a linear equation on $S_1(x)$:
\begin{gather}
    S_{1,\pm}= i\int\left(\mp\frac{\ve}{2p}+\frac{\vf\ve^2}{p^2}+\frac{\mu q_{\pm}}{2p^2}\right)d\vf,
\end{gather}
where $p$ is defined in Eq.~\ref{mom_semi}.
The integration gives $S_1(x)=-i\ln(\xi_1)$, $\xi_1$ is written in the main text (\ref{hbar1}).

\subsection{Validity of semiclassical expansion near the branch point of $p$.}
\label{app:branch1}
The validity of the semiclassical approach is checked as follows~\cite{LandauIII}.
We need to make sure that the 1$^{\rm st}$ order correction (in $\hbar$) to the l.h.s. of~\eqref{zero-order} is much smaller then the zeroth order expression.  

The first order in $\hbar$ term then reads:
\begin{gather}
    \label{expand1}
    \begin{split}
    &{\rm l.h.s.\ of\ \eqref{zero-order}}\Big|_{\rm 1^{st}\ order \ in\ \hbar}=-4i(\ve+\mu)\hbar \vf^\prime\\
    &\times
    \frac{p[\ve(\vf^2+1)-\mu]+2\ve^2\vf(\vf^2+1)-\mu^2\vf}{p(\vf^2+1)}
    \end{split}
\end{gather}
Now expand momentum $p$ near the branch point $p=0$ using the expansion of the field $\vf$:   $\vf(z)=i\sqrt{1-(\mu/\ve)^2}+i \delta z/a$. 
We see from the start that, unlike the denominator, the nominator of~\eqref{expand1} does not vanish as $p\rightarrow 0$. That means the first order semiclassical correction diverges at the branch point $p=0$. Plugging in $p=0$ in the nominator of ~\eqref{expand1} while keeping it finite in the denominator and setting $\vf^2+1=(\mu/\ve)^2$ we obtain the following:
\begin{gather}
    \label{expand2}
    \begin{split}
    &{\rm l.h.s.\ of\ \eqref{zero-order}}\Big|_{\rm 1^{st}\ order \ in\ \hbar}=\\
    &=2^{3/2}\hbar(\ve+\mu)(\ve^2-\mu^2)^{1/4}\sqrt{\frac{\ve}{a\delta z}}
    \end{split}
\end{gather}
Dividing it by the r.h.s. of ~\eqref{zero-order} (and squaring it for the sake of beauty) we obtain condition~\eqref{semi1} of applicability of the semiclassical expansion near the branch point of $p$ in the main body of the paper. 


\subsection{The correspondence between Schrödinger and Dirac equation}
\label{cooresp}
The goal of this paragraph is to show the full equivalence of semiclassical solutions following from the
Schroedinger equation~\eqref{eq_mod} and the main Dirac equation~\eqref{main}.
Expanding $\pi^2(x)$ formally in powers of $\hbar$ and keeping the first in $\hbar$ terms only we obtain:
\begin{gather}
    \pi(x) = \frac{1}{\hbar}\frac{p}{\vf^2+1}-\frac{i}{2p}\frac{\ve-\mu+\ve\vf^2}{(\vf^2+1)}\vf^\prime+...
\end{gather}
Next we write down the semiclassical solution in the standard way:
\begin{gather}
\label{sol2}
    \theta_{\pm}(x)=\frac{1}{\sqrt{\pi(x)}}\exp\left(\pm i\int^x \pi(x)\,dx\right)
\end{gather}
Finally, we perform the integration and express the semiclassical action in two convenient ways:
\begin{gather}
\label{mom2}
    \begin{split}
    \int^x\pi(x)\,dx&=-\frac{i}{2}\ln\frac{(p+\vf\ve)q_+\sqrt{\vf^2+1}}{\ve^2-\mu^2}\\
    &\equiv-\frac{i}{2}\ln\frac{\ve^2-\mu^2}{(p-\vf\ve)q_-\sqrt{\vf^2+1}}
    \end{split}
\end{gather}
Now, we make the expansion of $\eta(x)$ function up to the first in $\hbar$ terms:
\begin{gather}
    \frac{\eta(x)}{2}=\frac{i}{\hbar}\frac{\mu\vf}{\vf^2+1}+\frac{3}{2}\frac{\vf\vf^\prime}{\vf^2+1}
\end{gather}
and perform integration (in view of formula~\eqref{eq_mod2a}):
\begin{gather}
    \label{phase}
    \int^x\frac{\eta(x)\,dx}{2}=\frac{i}{\hbar}\int\frac{\mu\vf\,dx}{\vf^2+1}+\frac{3}{4}\ln(\vf^2+1)
\end{gather}
Combining two variants of action~\ref{mom2}, the integral~\eqref{phase} and solution~\eqref{sol2} and plugging the latter into wave function~\eqref{eq_mod2a} we obtain the exact correspondance with semiclassical expressions~\eqref{mom_semi}-\eqref{hbar1}.


\subsection{The expansion of the Schroedinger equation near the branch point}
\label{appendix:semi2}
In the vicinity of the branch point $z_\pm$ we have:
\begin{gather}
    \eta(\zeta)=-\frac{2}{\hbar}\frac{\ve\sqrt{\ve^2-\mu^2}}{\mu}+...
\end{gather}
 Now, the question is the expansion of $\pi^2(z)$ near $z_\pm$. It has the following shape:

\begin{gather}
\label{pi-exp}
    \pi^2(\zeta) = \underbrace{i\frac{(\ve-\mu)\ve^3}{\mu^3}\frac{1}{\hbar a}}_{\rm I} +\underbrace{
    \frac{1}{\hbar^2}\frac{2i\zeta}{a}\frac{\epsilon ^5 \sqrt{\ve^2-\mu ^2}}{\mu ^4}}_{\rm II}+...
\end{gather} 
Interestingly, once we take into account $z$ dependence of the potential $\vf(z)$ the potential stops vanishing at $z=z_\pm$. Let us prove that this is consistent with our analysis and this term can be discarded in comparison to the second one. We are interested in estimates, therefore we put $\ve\sim\mu$ everywhere in our arguments. Suppose  $\zeta$ is chosen in such a way that the first terms dominates over the second one:
\begin{gather}
\label{ratio1}
    \frac{\rm II}{\rm I} \sim \frac{z\ve}{\hbar}\ll 1 
\end{gather}
Then we have for the solution of ~\eqref{eq_mod} in the main body:
\begin{gather}
\label{estim1}
    \theta(\zeta)\sim \exp\left(\#\sqrt{\frac{\ve}{\hbar a}}\zeta\right).
\end{gather}
From the exponent of the~\eqref{estim1} we conclude that the characteristic scale of the wave function is $\sqrt{\hbar a/\ve}$.
\begin{gather}
    \frac{\rm II}{\rm I}=\sqrt{\frac{a\ve}{\hbar}}\gg1\ \hbox{(see Eq.~\ref{semi_cond})} 
\end{gather}
Therefore, in the region of interest (on the scale of a supposed semiclassical wave length) term I can be safely discarded comparing to II. 

Let us check that retaining just term II in~\eqref{pi-exp} is consistent with the analysis.
If  only term II is retained, the solution of~\eqref{eq_mod} in the semiclassical regime reads:
\begin{gather}
\label{estim2}
    \theta(\zeta)\propto \exp\left(i\int^\zeta\pi(\zeta)d\zeta\right)\sim \exp\left(\#\frac{\zeta^{3/2}}{\sqrt{a}}\frac{\ve}{\hbar}\right).
\end{gather}
As a result, the characteristic scale of the wave function is
$(a\hbar^2/\ve)^{1/3}$. Therefore, we have for the ratio:
\begin{gather}
    \frac{\rm II}{\rm I}=\left(\frac{a\ve}{\hbar}\right)^{1/3}\gg1 
\end{gather}
And indeed, this coincides with our initial statement. 
This way, term I in~\eqref{pi-exp} can be safely discarded, which concludes the validity of the semiclassical equation~\eqref{eq-mod2} in the main body of the paper.

\section{Analytic properties of Airy function asymptotics}
\label{app:airy}
To solve~\eqref{eq-mod2} we use the Laplace method of solution of differential equation with linear coefficients~\cite{goursat}. We use a concise exposition and notation from \cite{LandauIII}. To find the exact solution of the equation of the type:
\begin{gather}
\label{eq_init}
    \sum\limits_{m=0}^n(a_m+b_m s)\frac{d^m\theta}{ds^m}=0
\end{gather}
we compose the polynomials:
\begin{gather}
    P(t) = \sum\limits_{m=0}^n a_m t^m,\ \ Q(t) = \sum\limits_{m=0}^n b_m t^m, 
\end{gather}
And define the function:
\begin{gather}
    Z(t) = \frac{1}{Q}\exp\int\frac{P}{Q}\,dt
\end{gather}
determined up to a multiplicative factor.
Then the exact solution of~\eqref{eq-mod2} can be found in terms of a contour integral:
\begin{align}
    \theta(s) = \int\limits_C e^{s t}Z(t)\,dt
  \label{poly1}
\end{align}
where contour $C$ is chosen in such a way that function
\begin{gather}
\label{V_func}
    V = e^{st}QZ
\end{gather}
assumes identical values at its ends. 
The $V$ function controls the placement of the integration contour $C$ in the exact solution~\eqref{poly1}.

For the case of Airy equation:
\begin{gather}
    \theta^{\prime\prime}(s)+s\theta(s)=0
\end{gather}
the $P$ and $Q$ polynomials: 
\begin{gather}
\label{airy2}
    P=t^2,\ \ Q = 1 \ \Rightarrow\ Z(t)=e^{\frac{t^3}{3}}  
\end{gather}
Combining~\eqref{airy2} and~\eqref{poly1} we obtain the integral representation
\begin{gather}
\label{app:exact}
    \theta(s) = {\rm const}\int\limits_C e^{st+\frac{t^3}{3}}\,dt
\end{gather}
coinciding up to normalization constant with~\eqref{exact1}. 
Function $V$ is defined as:
\begin{gather}
\label{V1}
  V(t,s) = e^{st+\frac{t^3}{3}}  
\end{gather}
Now we need to choose the contour. Usually, the easiest points where $V$ assumes identical values to locate are the points where $V$ vanishes.  In case of a function as simple as ~\eqref{V1} it is enough to look at the large $t$ behavior of the exponent: $st+t^3/3\underset{t\rightarrow\infty}{\rightarrow}t^3/3$.  We see, that at large $t$ $V$ vanishes as long as 
\begin{align}
\cos(3\arg\,t)&<0,\ |t|\rightarrow\infty\Rightarrow\notag\\
\ \arg t&\in\left(\frac{\pi}{6}+\frac{2\pi n}{3},\ \frac{\pi}{2}+\frac{2\pi n}{3}\right)
\label{arg_t}
\end{align}
\begin{figure}[t]
	\centering
\includegraphics[width=1\columnwidth]{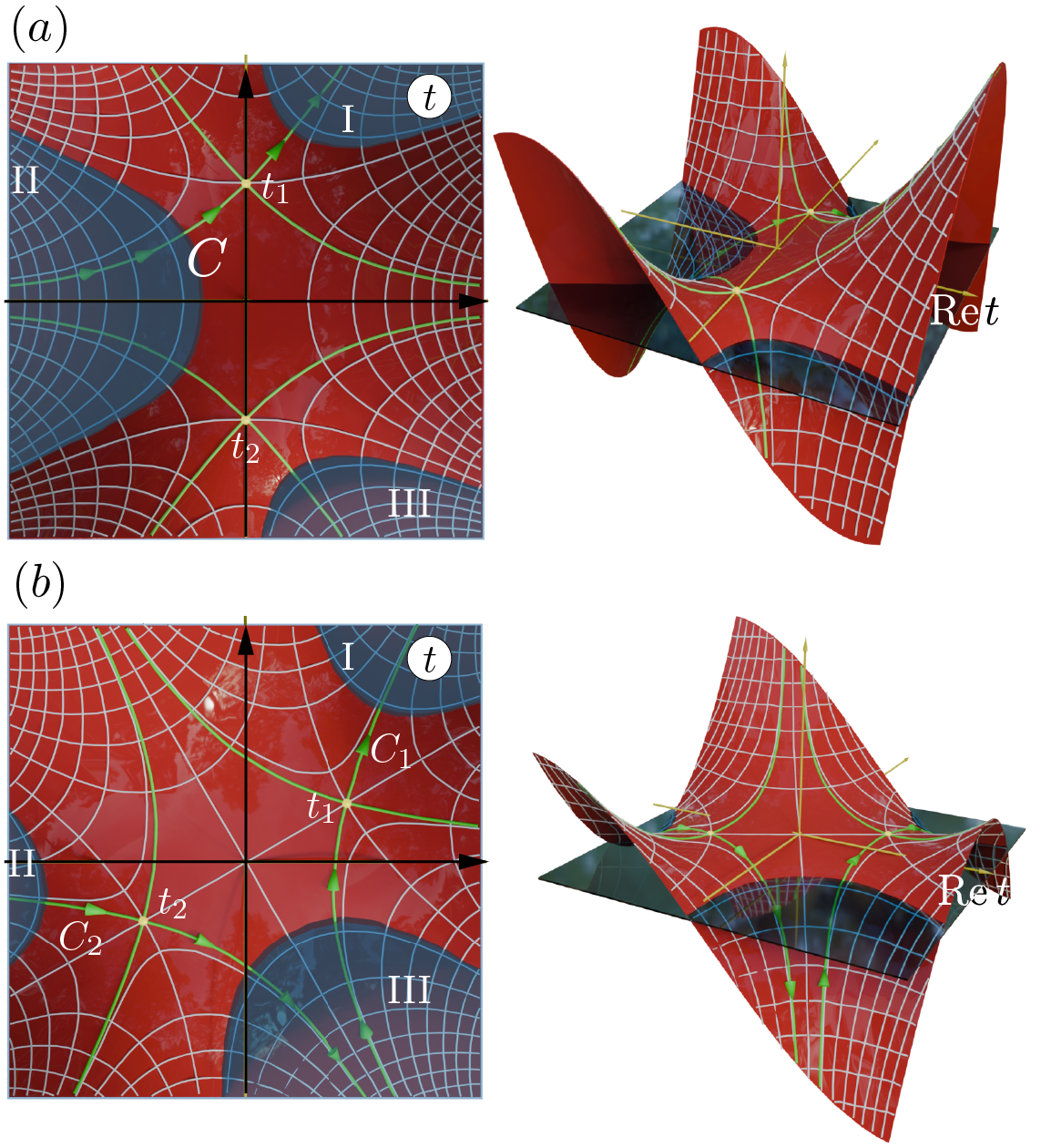}
	\caption{The relief of the real part of function $f(t) = st+t^3/3$ as a function of $t$. The grid is composed of the stationary lines of $f(t)$. The thin green contours are the steepest descent and ascent lines. They cross at two saddles (little yellow spheres). We also draw the horizontal plane for demonstrative purpose. The plane outlines the regions (I, II, III) where the value of ${\rm Re}\,f(t)$ tends to $-\infty$. These are the allowed regions for the end points of the placement of contour $C$ (function $V(s)\rightarrow0$ in these regions). These regions are also shown with gray color in Fig.~\ref{fig:asymp2}. (a) corresponds to $\arg\,s = 0$, (b) corresponds to $\arg\,s = -2\pi/3$.}
\label{fig:cont}
\end{figure}
Equation~\ref{arg_t} outlines the whole regions in the complex plane $t$ where contour $C$ should start and end. These regions are painted with gray color in Fig.~\ref{fig:asymp2} in the main part. They are also presented in Fig.~~\ref{fig:cont} with intersecting horizontal plane and denoted I, II and III for $n=1,2\ {\rm and}\ 3$ in Eq.~\ref{arg_t} respectively. It is important to note that regions~\eqref{arg_t} are independent of the arguments of complex variable $s$.  
\subsection{Asymptotics at $s\rightarrow+\infty$}
We need to place the contour in such a way that solution $\theta(s)$ has correct asymptotics at large positive value of $s$. That means, the integral~\eqref{app:exact} should be computed with the steepest descent method. To this end, we need to know possible stationary points of the exponent function of solution~\eqref{app:exact}:
\begin{gather}
\label{exp1}
    f(t) = st+\frac{t^3}{3}
\end{gather}
These are the saddle points:
\begin{gather}
    t_1 = i\sqrt{s},\ \ t_2 = -i\sqrt{s} 
\end{gather}
The second derivatives of function $f(t)$ at saddle points:
\begin{gather}
    f^{\prime\prime}(t_1) = 2i\sqrt{s},\quad f^{\prime\prime}(t_2) = -2i\sqrt{s}
\end{gather}
And the steepest descent directions at the saddles are given by the following relations:
\begin{align}
    \arg\,t &= \frac{\pi}{2}-\frac{\arg f^{\prime\prime}}{2}+\pi n,\ \ n\in\mathbb{Z}\ \Rightarrow
    \label{saddle_dir}\\
       \alpha_1& = \frac{\pi}{4}+\pi n,\quad \alpha = -\frac{\pi}{4}+\pi n  
\end{align}
The value of $f(t)$ at $t_1$ gives the main exponential value of solution~\eqref{app:exact} in the saddle point approximation: $\theta\propto \exp(2is^{3/2}/3)$ which matches the outgoing wave semiclassical asymptotics $\theta_+^{\rm app}$ from ~\eqref{asy1}. Therefore, contour $C$ in~\eqref{app:exact} should be placed in such a way that it may be deformed into the steepest descent path going through saddle $t_1$. This placement is depicted in Fig.~\ref{fig:cont}(a) and the saddle point approximation yields:
\begin{gather}
\label{steepest_airy1}
    \theta(s)={\rm const}\sqrt{\frac{2\pi}{|f^{\prime\prime}(t_1)|}}e^{f(t_1)+i\alpha_1} = 
    {\rm const}\frac{\sqrt{\pi}}{s^{1/4}}e^{\frac{2i}{3}s^{3/2}+\frac{i\pi}{4}}
\end{gather}
Comparing asymptotics ~\eqref{steepest_airy1} and semiclassical expression for $\theta_+^{\rm app}$ in ~\eqref{asy1} we easily fixate the constant in front of the integral:
\begin{gather}
\label{const}
    {\rm const} = \frac{e^{-i\pi/4}}{\sqrt{\pi}\gamma^{1/6}},
\end{gather}
arriving at normalized solution~\eqref{exact1}. 
\subsection{Asymptotics at $s\rightarrow \infty e^{-2\pi i/3}$}
After we rotate the argument of $s$ by $-2\pi/3$ to transfer to the respective anti-Stokes line the topography of the relief of ${\rm Re}\,f(t)$ changes.
The relief  ${\rm Re}f(t)$ and its steepest descent paths at $s=|s|\exp(-2\pi i/3)$ are presented in Fig.~\ref{fig:cont}(b). As we see, it is now possible to deform the initial  contour $C$ in such a way that it still begins in region II and ends in region I, yet it is split into two steepest descent paths $C_1$ and $C_2$ each of which passes via the respective saddle.  The steepest descent directions at saddles follow from condition~\eqref{saddle_dir}. 
\begin{gather}
    \alpha_1 = \frac{5\pi}{12},\quad \alpha_1 = -\frac{\pi}{12}
\end{gather}
And the contribution from two saddles reads:
\begin{gather}
\label{app:airy_fin}
    \theta(s)={\rm const}\frac{\sqrt{\pi}}{|s|^{1/4}}\left(e^{\frac{2i}{3}s^{3/2}+\frac{5\pi i}{12}}+ e^{-\frac{2i}{3}s^{3/2}-\frac{\pi i}{12}}\right)
\end{gather}
Asymptotics~\eqref{app:airy_fin} together with ~\eqref{const} gives formula~\eqref{stokes-left} in the main part of the paper.

\section{Exact equation}
\subsection{Transformation of the Dirac equation}
\label{app:trans}
Equation (\ref{main}) in the vicinity of the turning point $z_p$: $\vf(z)\approx ia/(z-z_p)$ and in the limit $|\zeta|=|z-z_p|\ll |a|$ becomes ($\hbar=1$):
\begin{gather}
\label{terrible1}
\begin{split}
4a\zeta&\Big[a\zeta\psi_1''\left(2 \zeta^2[\mu+\ve]-a\right)\\
&+\psi_1' \big(a ^2+4 \mu  \zeta^4 [\mu +\ve]
- 2 a  \zeta^2 [4 \mu +3 \ve ]\big)\Big]\\
&+\psi_1\Big[-3a^3+2a^2(7\mu+9\ve)\zeta^2+4a(\mu+\ve)(3\ve-5\mu)\zeta^4\\
&-8(\ve-\mu)(\mu+\ve)^2\zeta^6\Big]=0
\end{split}
\end{gather}
Surprisingly, Eq.~\ref{terrible1} can be solved in quadratures for $\mu\neq 0$ and in elementary functions for $\mu=0$.
Let us first study its asymptotics at $\zeta\rightarrow\infty$ and $\zeta\rightarrow 0$.
Retaining only the highest powers of $\zeta$ at $\zeta\rightarrow\infty$ we obtain the following equation:
\begin{align}
    a^2\psi''_1+2\zeta\mu a \psi'_1-\zeta^2(\ve^2-\mu^2)\psi_1 = 0\ \ \Longrightarrow\notag\\
    \psi_1 = e^{\frac{\zeta^2}{2a}(\ve\pm\mu)},\ \ \zeta\rightarrow\infty
    \label{as1}
\end{align}
At $\zeta\rightarrow0$ 
we retain the lowest powers of $\zeta$ to obtain:
\begin{gather}
    4u^2\psi''_1-4\zeta\psi'_1+3\psi_1=0,\ \Longrightarrow \psi_1 = \sqrt{\zeta}
\end{gather}
Finally, making a substitution  $\psi_1= e^{\frac{\zeta^2}{2a}(\ve-\mu)}\sqrt{\zeta}\psi(\zeta)$ we obtain a much simpler differential equation  ~\eqref{dif_simp}
in the main body.
\subsection{Asymptotics of the erf function}
\label{app:erf}
The ${\rm erf}\,(z)$ function has the Stokes line on the imaginary axis. That means, its asymptotic changes  as the argument of $z$ crosses $-\pi/2$ direction. Starting from the very definition of the erf function one easily derives the following asymptotics in the neightborhood of directions $-\pi/4$ and $-3\pi/4$:
\begin{gather}
\begin{split}
    {\rm erf}(\zeta)\Big|_{\zeta = |\zeta| e^{-i\pi/4}} &= 1-\frac{1}{\sqrt{\pi}}\frac{e^{-\zeta^2}}{\zeta}\\
    {\rm erf}(\zeta)\Big|_{\zeta = |\zeta| e^{-3i\pi/4}} &= 
    -1-\frac{1}{\sqrt{\pi}}\frac{e^{-\zeta^2}}{\zeta}
\end{split}
\end{gather}
Plugging in these asymptotics in to exact solution~\eqref{psi1exact} we immediately obtain relations~\eqref{const1} and~\eqref{asymp_pole} in the main body.

\section{Semiclassical functions in the vicinity of the turning point $z_p$}
\label{app:pole}
First, it is logical to compute the main exponential factors
~\eqref{action}, $S_{\pm}=\int_0^\zeta q_{\pm}(t)\,dt$. Here, the surprise awaits us, due to the subtlety of the work with a regular branch of momentum $p$ entering the definition of $q_{\pm}$.

To understand the behavior of $p$, we need to track the behavior of function $\vf^2(z)$. It has the second order pole at $z=z_p$:
\begin{gather}
\label{vf2}
    \vf^2(z) = -\frac{a^2}{(z-z_p)^2},\ \ z\rightarrow z_p.
\end{gather}
Looking at~\eqref{vf2} we see that there are two lines originating at the pole, along which $\vf^2(z)$ stays negative and real. Those are the lines with directions $\arg a$ and $\arg a+\pi$ going to the right and left respectively (for concreteness, we assume that $|\arg a|<\pi/2$):
\begin{align}
\label{lines}
    \zeta_{\rm right} &= |\zeta| \frac{a}{|a|}\quad\quad\quad\quad\ \, 
    \zeta_{\rm left} =  -|\zeta| \frac{a}{|a|},\\ \vf(\zeta_{\rm right}) &= i|\vf(\zeta_{\rm right})|\ \ \ 
    \vf(\zeta_{\rm left}) = -i|\vf(\zeta_{\rm left})|
    \label{vf}
\end{align}
By the very definition of these lines they are the steepest ascent paths of function ${\rm Re}\vf^2$.  ${\rm Re}\vf^2(z)$ changes from $-\infty$ to $0$ as ${\rm Re}z\rightarrow\pm\infty$.
To make things more transparent, we draw these lines for the Lorentzian function $\vf(z) = (z^2+1)^{-1}$ in Fig. ~\ref{fig:branch2} (They flow below both branch cuts starting at $z_\pm$). 

As point $z$ moves from the pole $z_p$ to the right along the steepest ascent line, it inevitably hits the branch point of $p$, $z_+$, which should be then circumvented from below. The same happens as the point moves to the left (point $z_-$ is hit). 
The importance of these lines, therefore, rest in the fact that squared momentum $p^2$ stays real and positive to the right of $z_+$ and negative to the left (and vice versa for $z_-$).  Therefore, the regular branch of $p$ changes its sign as $x$ moves from $+\infty$ to $-\infty$.

The definition of the regular branch of $p$, presented right after Eq.~\ref{mom_semi} in the main body, dictates its value once point $z_+$ is passed from right to left along the lower semicircle: 
\begin{gather}
    p(\zeta_{\rm right}) = -i|p| \equiv e^{-i\pi/2}\sqrt{\ve^2|\vf^2|-\ve^2+\mu^2}
\end{gather}
Now we Taylor-expand the last equation in the vicinity of $z_p$ while sticking to the right steepest ascent line:
\begin{align}
\label{pleft}
    p(\zeta_{\rm right}) &= 
    e^{-i\pi/2}\left(\ve|\vf|-\frac{\ve^2-\mu^2}{2\ve|\vf|}\right)+...\\
\label{qplus}
    q_{+}(\zeta_{\rm right})&=e^{i\pi/2}\frac{\ve+\mu}{|\vf|}
\end{align}
Using~\eqref{qplus} we are ready to compute the main exponential factor $S_+$:
\begin{gather}
    S_+(\zeta_{\rm right}) = e^{i\pi/2}\int_0^{\zeta_{\rm right}} \frac{\ve+\mu}{|\vf|}d\zeta_{\rm right},
\end{gather}
where the integration is assumed to be done along the right steepest ascent line.
Next, we change according to ~\eqref{lines}:
\begin{gather}
\label{cont}
    |\vf| = \frac{|a|}{|\zeta|} = \frac{a}{\zeta_{\rm right}}\rightarrow\frac{a}{\zeta}.    
\end{gather}
The last equality in~\eqref{cont}
 is the analytical continuation from the right steepest ascent line to its neighborhood. 
 We obtain:
\begin{gather}
\label{splus}
    S_{+,>} = i\int_0^\zeta \zeta\frac{\ve+\mu}{a}d\zeta = i(\ve+\mu)\frac{\zeta^2}{2}.
\end{gather}
Now we need to compute pre-exponential factors $\xi_{1,\pm}$ (Eq.~\ref{hbar1}).
In fact, we are \textit{almost} ready to extract the correct regular branch of the square root $\xi_{1,+}$. 

Surprisingly, function $p+\vf \ve$ entering the nominator of the expression under the square root of $\xi_{1,+}$ never vanishes in the complex plane. We assume its argument to be zero  at $x\rightarrow+\infty$. One easily convinces oneself that as we arrive in the neighborhood of $z_p$ along the right steepest ascent line, the argument of $p+\vf\ve$ becomes $\pi/2$. This is also clearly seen from eq.~\ref{pleft}:
\begin{gather}
\label{aux}
    \ve\vf+p\Big|_{\rm right} = e^{i\pi/2}\frac{\ve^2-\mu^2}{2\ve|\vf|}
\end{gather}
Collecting~\eqref{pleft},~\eqref{qplus} and~\eqref{aux} we obtain for $\xi_{1,+}$:
\begin{gather}
    \xi_{1,+}\Big|_{\rm right}=
    e^{3\pi i/4}\frac{\ve+\mu}{\ve}\sqrt{\frac{\ve-\mu}{2|\vf|^3}}+...
\end{gather}
Making, as before, the analytical continuation $\zeta_{\rm right} = \zeta$ we obtain:
\begin{gather}
\label{xiplus}
    \xi_{1+,>}=
    e^{3\pi i/4}\frac{\ve+\mu}{\ve}\sqrt{\frac{\ve-\mu}{2}}\left(\frac{a}{\zeta}\right)^{3/2}
\end{gather}
Collecting~\eqref{xiplus} and~\eqref{splus} we obtain relation for $\psi_{1+,>}$ in~\eqref{qctpasympt} in the main body.

In the same manner we obtain relation for $S_-$ and
the rest of $\xi_1$:
\begin{figure}[h!]
	\centering
	\includegraphics[width=0.75\columnwidth]{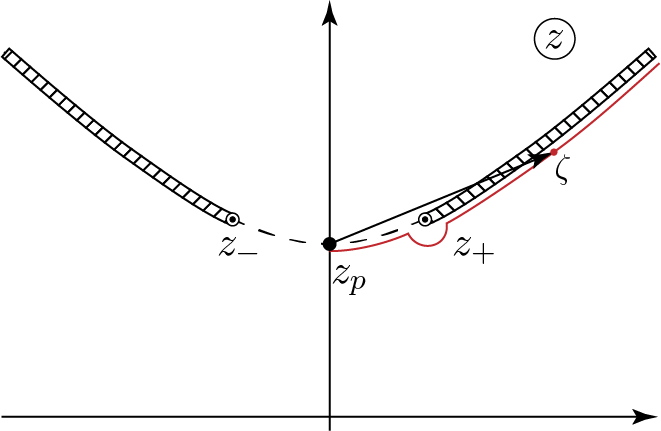}
	\caption{The curves ${\rm Im}\vf^2(z) = 0$ for the Lorentzian function $\vf(z)=(z^2+1)^{-1}$. The curves form hyperbola $y=\sqrt{x^2+1}$.}
\label{fig:branch2}
\end{figure}
\begin{gather}
\label{app_an}
    \begin{split}
        \xi_{1+,<}(x) &= e^{-3\pi i/4}\frac{\ve+\mu}{\ve}\left[\frac{\ve-\mu}{2}\right]^{1/2}\frac{1}{|\vf|^{3/2}},
        \\
        \xi_{1-,\gtrless}(x) &= e^{\pm i\pi/4}\left[\frac{2(\ve-\mu)}{|\vf|}\right]^{1/2}    
    \end{split}
\end{gather}
Performing analytical continuation in~\eqref{app_an} we get the rest of the asymptotic formula \eqref{qctpasympt}.

\section{Hamiltonian transformation and Born scattering}
\subsection{Derivation of the unitary transform $\hat{U}(x)$.}
\label{app:trans}
In the absence of magnetic field Hamiltonian reads:
\begin{gather}
H=v_F\sigma_y\hat p+\frac{\sigma_z}{2}(\vf\hat p+\hat p\vf)
\end{gather}
We introduce the following unitary transformation:
$\hat{U}(x)=\exp (i\theta(x)\sigma_x)$, 
where $\theta (x)$ is real for real $x$. The next step is to apply this transformation to $H$. The transformed Hamiltonian $\tilde{H}=\hat U^\dag H \hat U$ can be expressed as
\begin{gather}
\label{ham_trans}
\tilde{H}\!=\!v_F(\hat U^\dag \sigma_y \hat U)(\hat U^\dag \hat p \hat U)
\!+\!\frac12(\vf\hat U^\dag \hat p \hat U+\hat U^\dag \hat p \hat U\vf)(\hat U^\dag \hat \sigma_z \hat U).
\end{gather}
Let us write down the transformations of the individual terms: $\hat U^\dag \hat p \hat U=\hat p+\theta'\sigma_x$, $\hat U^\dag \sigma_y \hat U=(\sigma_y\cos 2\theta+\sigma_z\sin 2\theta)$, $\hat U^\dag \sigma_z \hat U
=(\sigma_z\cos 2\theta-\sigma_y\sin 2\theta)$.
One plugs in these expressions into~\eqref{ham_trans}. 

Our goal is to find such function $\theta(x)$ that the term proportional to $\sigma_y$ in the transformed Hamiltonian vanishes. 
We immediately obtain two consistent equations:
\begin{gather}
\begin{split}
\sin2\theta-\vf\cos2\theta & = 0\\
\frac{1}{2}\frac{d}{dx}[\sin2\theta-\vf\cos2\theta] & = 0
\end{split}
\end{gather}
This way we recover identity~\eqref{theta} and Hamiltonian~\eqref{ham_exact} as well as~\eqref{pot_transf} in the main body of the paper.
\subsection{Exact 
eigenfunctions of the unperturbed Hamiltonian}
\label{app:eigen}
The eigenfunctions of Hamiltonian~\eqref{ham_exact} are needed for further perturbative analysis and can be easily found. The corresponding equation reads:
\begin{gather}
    -i\left[v(x)\psi_{1,2}^\prime(x)+\frac{1}{2}\psi_{1,2}(x)v^\prime(x)\right] = \pm\ve\psi_{1,2}(x)
\end{gather}
The l.h.s can be simplified with integrating factor as follows:
\begin{gather}
    -i\sqrt{v}\left[\sqrt{v}\psi_{1,2}^\prime+\frac{v^\prime}{2\sqrt{v}}\psi_{1,2}\right] = -i\sqrt{v}[\sqrt{v}\psi_{1,2}]^\prime
\end{gather}
And we trivially obtain eigenfunctions~\eqref{sol_exact} in the main body.
The important property of the eigenfunctions is the correct completeness relation. We obtain:
\begin{gather}
\label{unity_res}
\sum\limits_\sigma\int\limits_{-\infty}^{\infty}\underset{\sigma}{\psi_\ve}(x)\underset{\sigma}{\psi_\ve}^\dag(x) \frac{d\ve}{2\pi}= \mathbf{1}\delta(x-x^\prime)
\end{gather}
where $\sigma = \leftrightarrows$ and $\mathbf{1}$ is the 2D unit matrix.
Therefore, the correct measure counting the quantum eigenstates is $d\ve/(2\pi)$.
Also, of note, the normalization condition imposed on eigenfunctions. We need to make sure that condition
\begin{figure}[h!]
	\centering
\includegraphics[width=0.6\columnwidth]{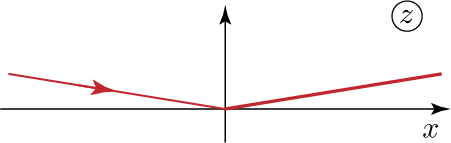}
	\caption{Deformation of the contour of the integral~\eqref{norm}}
\label{fig:conta}
\end{figure}
\begin{gather}
\label{norm}
    \int\limits_{-\infty}^\infty\underset{\sigma}{\psi_\ve}^\dag(x)
    \underset{\sigma}{\psi_{\ve^\prime}}(x)\,dx = \int\limits_{-\infty}^\infty e^{\pm i(\ve-\ve^\prime)\tau(x)}\frac{dx}{v(x)}\propto\delta(\ve-\ve^\prime)
\end{gather}
is satisfied.
Making a change $x\rightarrow\tau(x)$ we arrive at the integral
\begin{gather}
    \int\limits_{-\infty}^\infty e^{\pm i (\ve-\ve^\prime)\tau}d\tau = 2\pi\delta(\ve-\ve^\prime)
\end{gather}
Strictly speaking at $\ve\neq\ve^\prime$, the integral~\eqref{norm} converges only in the case when the contour is bent upward (or downward depending on the sign of coefficient in front of $\tau(x)$ in the exponent) in the complex plane, as shown in Fig.~\ref{fig:conta}. 

\subsection{The Green's function}
\label{green}
To build the perturbation theory we need a system's Green's function.
We define the retarded Green's function as inverse Schrodinger operator, $\hat{G}=(\epsilon-\hat{H}+i0)^{-1}$. In the basis of the eigenfunctions it is expressed as
\begin{gather}
G(\epsilon;x,x')=\sum\limits_{\alpha}
\frac{\psi_\alpha(x)\psi^\dag_\alpha(x^\prime)}{\epsilon-\ve_\alpha+i0}
\end{gather}
Substituting eigenfunctions~\eqref{sol_exact} and using the resolution of unity~\eqref{unity_res}, we get
\begin{align}
G(\ve;x,x')=\int\frac{d\ve'}{2\pi}
\frac{\ve+\ve'\sigma_z}{\sqrt{v(x)v(x')}}
\frac{e^{i\ve'(\tau(x)-\tau(x'))}}{(\ve+i0)^2-\ve'^2}\label{green_fun}
\end{align}
Integrating the last expression with the help of residue theorem we obtain the Green's function~\eqref{green} in the main text. 

\section{The analytical properties of $\tau(z)$}
\label{app:analyt}
\begin{figure}[t!]
	\centering
	\includegraphics[width=1\columnwidth]{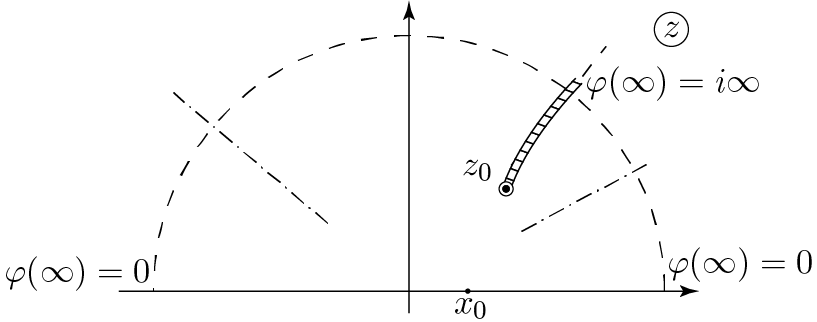}
	\caption{The schematic behavior of the potential function $\vf(z)$ and $\tau(z)$. The dashed radian lines denote the boundaries of the change of the limit of $\vf(\infty)$.}
\label{fig:essential}
\end{figure}
We are going to argue that in general situation, there is a stationary ${\rm Re}\,\tau(z) = {\rm const}$ path connecting the real axis and the branch point $z_0$ of function $\tau(z)$.

Before proceeding further, let us make the following comment. The stationary path of a function of complex variable stemming from a particular point has 3 possible placements: i) it forms a closed contour returning to the starting point. ii) it ends in the singularity of the function (a pole,  an essential singularity) iii) it has a corner at a branch point or at the stationary point. In particular, it turns by $\pi$ and returns to the initial point in the case of the branch point of the second order (square root).  

\subsubsection{The behavior of function $\vf(z)$ at $z\rightarrow\infty$}
Since $\vf(\pm\infty) = 0$ the function $\vf(z)$ can't have a polynomial behavior at $z\rightarrow\infty$.
Therefore, its Laurent series in the vicinity of $\infty$ has infinite amount of terms with positive powers of $z$ and $z=\infty$ is an essential singularity.

According to Casorati-Weierstrass theorem, function $\vf(z)$ can have any limiting value depending on the direction at which $z\rightarrow\infty$.  
We understand that due to the continuity of $\vf(z)$, there is a sector (denoted by dash-dotted lines in Fig.~\ref{fig:essential}) where $\vf(\infty)=0$. On the other hand, we may always find the direction $\arg z>0,\ |z|\rightarrow\infty$ such that directional limit of $\vf(|z|e^{i\arg\,z})\rightarrow i\infty$ (dotted line in Fig.~\ref{fig:essential}). 

\subsubsection{On the placement of the branch cut}
Next, let us draw a branch cut of function $\tau(z)$ stemming from point $z_0$ upwards  (away from the real axis) in the direction, where ${\rm Im}\,\vf^2(z) = 0$. It is the steepest descent direction of ${\rm Re}\,\vf^2(z)$ and precisely the direction of the dashed line in Fig.~\ref{fig:essential} discussed before. Therefore, $\vf^2(z)+1$ is real and negative and the square root $\sqrt{\vf^2(z)+1}$ is purely imaginary.
That entails, in particular, that on both sides of the branch cut:
\begin{gather}
    \label{aux1}
    \tau(\infty\pm\delta) = \tau(z_0)\pm\frac{1}{i}
    \int\limits_{z_0}^{\infty\pm\delta}
    \frac{dx+i dy}{\sqrt{|\vf^2(z)-1|}}.
\end{gather}
where $\infty\pm\delta$ denotes the infinite points below (above) the branch cut.
It is important to note that the direction of the branch cut is chosen in such a way that  $\vf^2(z)\rightarrow\infty$. Therefore, the branch cut at $z\rightarrow\infty$ lies above the limiting separatrix (see Fig.~\ref{fig:essential}). Reading formula~\eqref{aux1} we point out that in the general situation
\begin{gather}
    \frac{dy}{\sqrt{|\vf^2(z)-1|}}\neq0.
\end{gather}
The last expression then entails:
\begin{gather}
\label{neq}
    {\rm Re}\,\tau(\infty\pm\delta)\neq {\rm Re}\,\tau(z_0) 
\end{gather}
It is also important to note that the integral entering~\eqref{aux1} can in principle be convergent.

\subsubsection{The stationary path ${\rm Re}\,\tau(z) = {\rm Re}\,\tau(z_0) $}
We notice that the real axis is the stationary path for ${\rm Im}\,\tau$: ${\rm Im}\,\tau = 0$. Therefore, all the stationary paths of ${\rm Re}\,\tau(z)$ stems vertically upwards and downwards from the real axis. One immediately convinces oneself that the upward direction is the steepest descent path for ${\rm Im}[\tau(z)]$. As was argued before, function $\vf(z)$ ought to have an essential singularity at infinity, therefore, function $\tau(z)$ also has an essential singularity at $z\rightarrow\infty$. 
Since $\vf(z)$ has no poles, function $\tau(z)$ has no stationary points. 

That means,  the stationary paths of ${\rm Re}\,\tau(z)$ stemming upwards from the real axis have only two options: they go to infinity, or they end at the second order branch point $z_0$ (and return back) to the starting point on the real axis.

Now we are going to argue that there always exists a point on a real axis with the value of $\tau(x)$ equal to the real part of $\tau(z_0)$ at the branch point. Indeed, the right real semi line is the line of the steepest ascent of function $\tau(z)$ (In fact, $\tau(\pm\infty)=\pm\infty$). Therefore, it contains all possible values   ${\rm Re}[\tau(z)]$ may assume in the complex plane. Consequently, there is a point $x_0$ on the real axis where
$\tau(x_0) = {\rm Re}[\tau(z_0)]$. On the other hand, if the steepest descent curve stemming from $x_0$ doesn't enter $z_0$, it should go to infinity. The latter  means that  points with identical values of ${\rm Re}\,\tau(\infty) = {\rm Re}\,\tau(z_0)$. However, this is not possible, due to relation~\eqref{neq}. This way, we argued that there is always the steepest descent path stemming from some point $x_0$ on the real axis and ending at point $z_0$.

\subsubsection{Other stationary paths}
As was pointed out in the previous paragraph,
function $\tau(z)$ has different limits at $z=\infty$ below and above branch cut: ${\rm Re}\tau_{\infty\downarrow},\ {\rm Re}\tau_{\infty\uparrow}$. Finding the points on the real axis with the same values of $\tau$ (the latter are real) we can draw the steepest descent lines, as shown in Fig.~\ref{fig:deform_init}.

\section{On the meaning of anti-Stokes and Stokes lines}
\label{app:asymp}

The importance of anti-Stokes lines lies in the fact that these are the lines along which both linear independent solutions of the differential equation neither grow nor get suppressed exponentially. 
The exact solution in our case is always represented by some contour integral $\psi(\zeta)\propto\int e^{\zeta f(t)}g(t)\,dt$, i.e. Airy or erf integral in previous sections or a Bessel integral (see below). The asymptotics  of the solution is then given by the steepest descent paths going through the stationary point (or round singular points) of  ${\rm Re}\,f(t)$ in terms of the asymptotic expansion near the latter point.  

As one changes the argument $\zeta$, the relief of exponential function ${\rm Re}\,[\zeta f(t)]$ gets deformed.
The deformation of the mentioned relief changes the steepest descent path, bringing addition saddles (or singularities) of $f(t)$ into its vicinity.  The scattered wave appears as an additional asymptotic series at the new saddle (or singularity) in the steepest descent path. 

If $\zeta$ belongs to the anti-Stokes line, neither of contributions (yielding incident and scattered waves) exponentially dominate.
Once we deviate from the anti-Stokes line, the additional contribution giving the scattered wave becomes exponentially suppressed comparing to the leading contribution. The main contribution, however, has an inherent error built in any asymptotic series. The question one asks: what if the error of the leading asymptotic series is actually larger than the exponentially small contribution yielding the scattered wave?

Fortunately, the  asymptotic analysis has an answer to this concern. The statement is as follows: the error in the \textit{optimally summed} dominant asymptotic series is always smaller than the subdominant series, except for so-called Stokes lines (the lines in the complex plane where the leading contribution maximally dominates over the subleading one).
To perform the optimal summation of the asymptotic series one needs to deform the integration contour representing the solution along the \textit{global} steepest descent path of the function ${\rm Re}\,\zeta f(t)$~\cite{Olver}. 

\section{Semiclassical functions near the branch point $p$ at $\mu\ll\ve$}
\label{app:semi_mu}
\begin{figure}[t]
	\centering
\includegraphics[width=0.45\columnwidth]{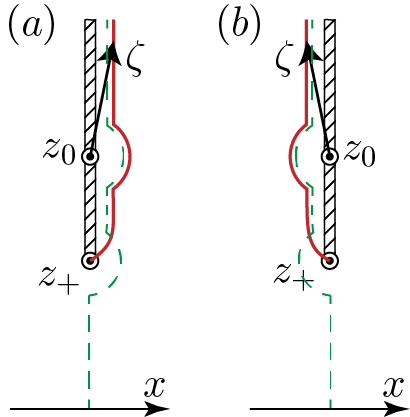}
	\caption{The integration contours (red curves) along the (a) right anti-Stokes line, (b) left anti-Stokes line. The green dashed lines represents the path along which the argument of $p$ follows as point $\zeta$ travels from the real axis, where $p$ is real and positive, upward along the right(a) and left (b) anti-Stokes lines}
\label{fig:contb}
\end{figure}
We start from the computation of $q_+$. Plugging in expansion~\eqref{pole2} into expression~\eqref{hbar1} and retaining the two leading terms we have:
\begin{gather}
\label{qpm}
    q_{\pm}(\zeta) = \frac{-\mu i+\ve\sqrt{\frac{2i\zeta}{a}}}{\frac{2i\zeta}{a}}+...
\end{gather}
Next, we integrate in the direction $\zeta = iy$ from the point $z_+$ along the right bank of the branch cut starting at point $z_+$ (the right anti-Stokes line, see Fig.~\ref{fig:contb}(a)). 
The relative (with respect to $z_0$) coordinate of $z_+$ is found via the Taylor expansion:
\begin{align}
    \vf(z)=i+\frac{\zeta}{a}+...=i\sqrt{1-\frac{\mu^2}{\ve^2}}=i-\frac{i\mu^2}{2\ve^2}+...\notag\\
    \Rightarrow\zeta\equiv\zeta_+=-\frac{ia}{2}\frac{\mu^2}{\ve^2}+...\label{zplus}
\end{align}
Now it is time to compute the regular branch of the square root (which is nothing but momentum $p = \ve\sqrt{2i\zeta/a}$) entering $q_+$ in~\eqref{qpm}. As seen from Fig.~\ref{fig:contb}(a) (green dashed line), the argument of $p$ rotates by $\pi$ as $\zeta$ travels from real axis to the right anti-Stokes line. Therefore, $p = i|p|$ and the square root in~\eqref{qpm} is expended as:
$\sqrt{2i\zeta/a} \equiv \sqrt{-2y/a} = i\sqrt{2y/a}$. Consequently, the semiclassical action reads:
\begin{gather}
\label{action0}
    S_{+} = -\frac{\mu a}{2}\int\limits_{\zeta_+}^{iy}\frac{d\zeta}{\zeta}+\ve\int_0^y\sqrt{\frac{a}{2y}}\,dy
\end{gather}
where  we changed the lower integration limit in the second integral from $y=-i\zeta_+$ to zero, since the upper integration limit obeys the condition $|y|\gg|\zeta_+|$ according to ~\eqref{semi2}. This approximation, however, cannot be done with the first integral, since it becomes log divergent at $\zeta\rightarrow0$. The first integral in~\eqref{action0} is logarithmic, and we need to track correctly the change of the argument of $\zeta$. It is easily red from Fig.~\ref{fig:contb}(a): $\Delta\arg\zeta = \pi$. Finally, we obtain
\begin{gather}
\label{action_semi}
    S_{+} = -\frac{\mu a}{2}\ln\frac{2y\ve^2}{a\mu^2}-\frac{i\pi\mu a }{2}+\ve\sqrt{2ay}
\end{gather}
Now we are ready to tackle the pre-exponential factors $\xi_{1,\pm}$ in~\eqref{hbar1}. It is understood that $|\vf\ve/p|\sim\sqrt{|a\zeta|}\gg1$. Therefore, one can discard $1$ under the square root in the definition of $q_\pm$. We already argued above in this section that the argument of $p$ is $\pi/2$. The same is true for $\vf = i$ in the vicinity of $z_0$. As a result $\arg[\ve\vf/p] = 0$. On the other hand we can discard term with $\mu$ in the zeroth order approximation in~\eqref{qpm}. Therefore, $\arg q_+ = \arg p-\arg(\vf^2+1) = \pi/2-\pi = -\pi/2$. Finally, we have
\begin{gather}
\label{xiplusfin}
    \xi_{1,+} = |\xi_{1,+}|e^{-\frac{i\pi}{2}} = \frac{\ve a}{y}e^{-\frac{i\pi}{2}}.
\end{gather}
Combining~\eqref{xiplusfin}, ~\eqref{action_semi} and plugging them into semiclassical wave function~\eqref{hbar1} we obtain $\psi_{+,>}$ (Eq.~\ref{sem_reg1}) in the main part of the paper. In complete analogy, one obtains relations~\eqref{sem_reg2} and~\eqref{sem_reg3} as well.

\section{Exact solution near the branch point $p$ at $\mu\ll\ve$}
\label{app:fin}
To solve~\eqref{bessel1} we use the Laplace method of solution of differential equation with linear coefficients outlined in Appendix~\ref{app:airy}.
Polynomials $P$ and $Q$ are red from equation~\eqref{bessel1}:
\begin{gather}
    P = (3i-2a\mu)t+\ve^2a,\quad Q=2it^2.
\end{gather}
Performing integration $\int (P/Q)\,dt$ and changing $t = \ve s$ we obtain the solution in the form~\eqref{exact2} (up to a contant in front of the integral). Function $V(\zeta, t)$ from~\eqref{V_func} becomes:
\begin{gather}
    V(\zeta, s) = \exp\left(\zeta\ve s+\frac{\ve a i}{2s}\right)s^{ia\mu+\frac{3}{2}}.
\end{gather}
To define the regular
branches of the multivalued function $s^{ia\mu}$ entering the exact solution and function $V$, a branch cut should be drawn from the  point $s=0$.
The most suitable direction is upward.

\subsection{The placement of the contour}
We are looking for points in the complex plane $s$ where function $V(\zeta, s)$ assumes identical values. The points which are the easiest to identify are the ones where $V$ vanishes.

Suppose $\zeta$ is initially placed on the right anti-Stokes line (Fig.~\ref{fig:deform3}(a) right).
As in the main body, let us assume without loss of generality $a$ to be real.
We see that if $s\rightarrow+i\infty$ function $V(\zeta, s)\rightarrow 0$.
On the other hand, if $s\rightarrow 0$ in  vertical direction (see Fig~\ref{fig:deform3}(a) left) $V$ vanishes as well. 
Therefore, contour $C$ depicted in Fig.~\ref{fig:deform3}(a) satisfies the principal condition of the placement.
\subsection{Saddle point approximation, right anti-Stokes line}
To match the exact solution with semiclassical expressions, we need to find the asymptotics of~\eqref{exact2} at large $\zeta$.
As pointed out in the main body of the paper, the exponent function~\eqref{func} has two saddles $s_{1,2}$, the right being the one giving the correct semiclassical exponential~\eqref{sem_reg1}. 
The steepest descent directions at the saddles are given by Eq.~\ref{saddle_dir} and read:
\begin{align}
\label{app:steepest}
    \alpha(s_{1,2}) = \pm\frac{\pi}{4}+\pi n,\ n\in\mathbb{Z}
\end{align}
Computing the integral in the saddle point approximation and taking into account that the steepest descent direction at $s_1$ is $\pi/4$ and comparing the result with~\eqref{sem_reg1} we are able to fixate the constant in front of the integral in~\eqref{exact2}. 

\subsection{Saddle point approximation, left anti-Stokes line}
Now we need to build an analytical continuation of the exact solution given by integral~\eqref{exact2} once $\zeta$ goes from the right to the left anti-Stokes line and rotates by $2\pi$ in the clockwise direction.

Contour $C$ should guarantee the vanishing of $V(\zeta, s)$ at $s\rightarrow\infty$ during all the transformation of $\zeta$. 
To compensate for the change of argument of $\zeta$ by $-2\pi$, the end point of the contour should rotate by $2\pi$ in the complex plane of $s$ (as depicted in Fig.~\ref{fig:deform3}(c)) turning it into a spiral. Since the branch cut obstructs the rotation of the contour, the latter should cross the branch cut and continue on the second Riemann sheet of multivalued function $s^{ia\mu-1/2}$.
All the arguments of $s$ on the second Riemann sheet are related to the ones on the first one by  $2\pi$ rotation: $s|_{\rm second} = s|_{\rm first} e^{2\pi i}$.

The steepest descent paths of function $f$~\eqref{func} are the same on both Riemann sheets. Therefore, to find the asymptotics of the integral after the transformation $\zeta\rightarrow\zeta e^{2\pi i}$,  the contour is deformed  along the steepest descent curves on both Riemann sheets, as shown in Fig.~\ref{fig:principal3}. The arc of the contour at $\infty$ (Fig.~\ref{fig:principal3}(a)) does not contribute to the integral, since the integrand vanishes at all points of the arc.

From the structure of the contour in Fig.~\eqref{fig:principal3} we see that both saddles now contribute to the integral. Saddle $s_2$ is passed in $-\pi/4$ direction, while saddle $s_1$ is passed twice. The contribution of the saddle from the second Riemann sheet is identical to the one from the first Riemann sheet up to a constant factor coming from the different value of the multivalued function at saddle $s_1$:
\begin{gather}
    s_1^{i\mu a-1/2}\rightarrow s_1^{i\mu a-1/2} e^{2\pi i(i\mu a-1/2)}.
\end{gather}
The last equation leads directly  to~\eqref{victory2} after simple algebra.


\bibliographystyle{apsrevlong_no_issn_url}

\bibliography{ti.bib}
\end{document}